\documentclass[12pt]{article}
\usepackage{amssymb,amsmath}
\usepackage{cite}
\usepackage{arydshln}

\newtheorem{lemma}{Lemma}
\newtheorem{prop}{Proposition}
\newtheorem{thm}{Theorem}
\newtheorem{cor}{Corollary}

\def\id{\mbox{id}}

\begin{document}

\begin{center}
{\large\bf Invariants and matrix elements of the quantum group $U_q[gl(n,\mathbb{C})]$ revisited}\\
~~\\

{\large Mark D. Gould and Phillip S. Isaac}\\
~~\\

School of Mathematics and Physics, The University of Queensland, St Lucia QLD 4072, Australia.
\end{center}

\begin{abstract}
In a previous paper the generator matrix elements and (dual) vector reduced Wigner coefficients
(RWCs) were
evaluated via the polynomial identities satisfied by a certain matrix constructed from the
$R$-matrix $R$ and its twisted counterpart $R^T=T\circ R$. Here we provide an alternative evaluation
utilising the $R$-matrix $\tilde{R} = (R^T)^{-1}$. This provides a new direct derivation of the vector
RWCs obtained indirectly in earlier work via a symmetry relation. This approach
has the advantage that it generalises to the Lie superalgebra case, which will
be investigated elsewhere.
\end{abstract}

\section{Introduction}

Quantum groups were introduced in the mid-1980s as deformations of the
universal enveloping algebra of Lie algebras \cite{J85,D86}, mainly due to
their relevance to the quantum Yang-Baxter equation, a key ingredient to the
quantum inverse scattering method \cite{STF79}.
Crucial to their applicability to such problems of physical interest, is a
well-developed, and {\em explicit}, theory of representations. 
The key is to have access {\em on demand} to explicit
expressions for quantities such as Wigner coefficients and matrix elements, ready to be used for
applications  such as the analysis of physically motivated quantum integrable systems.

Such a constructive approach was applied first to Lie algebras by Baird and
Biedenharn \cite{BB1963} using pattern calculus methods, motivated by the
seminal results of Gelfand and Tsetlin \cite{GT1950}. 
Then, using the characteristic identities originally developed systematically by Green and
Bracken \cite{Green1971,BraGre1971}, in a series of papers Gould presented a
thorough derivation of various quantities such as formulae for reduced Wigner coefficients
(isoscalar factors), Wigner coefficients, reduced matrix elements and matrix
elements 
\cite{Gould1978,Gould1980,Gould1981,Gould1981b,Gould1986,Gould1986b}. It is worth
pointing out that for the case of simple Lie algebras, the form of the
characteristic identities depend on the quadratic Casimir element.

The approach using characteristic identities was then further
developed for quantum groups \cite{GLB1992,Gould1992}. In the case of Lie
algebras, several identities are required to ultimately derive the relevant
formulae -- one for the characteristic matrix $A$, and another for the
so-called adjoint characteristic matrix,
$\overline{A}$. In the case of $gl(n)$ for instance, both matrices partition into a convenient
block form that highlights the subalgebra embedding $gl(n-1)\subset gl(n)$. For
the quantum group case, the form of the characteristic identities do
still depend on the Casimir element, which in turn is expressible in terms of
the universal $R$-matrix. Furthermore, only one of the characteristic matrices
partitions conveniently (depending on the choice of $R$-matrix), but this does
not affect the calculation, as symmetry relations involving $q$-dimensions of
the irreducible representations exist among the
invariants associated to the two identities. These symmetry relations may be employed 
to streamline the derivation (see, for example, equation (38) in \cite{GLB1992}) and bypass the
issue of partitioning.

Along these lines, more recent work by the current authors have seen a
development of characteristic identities utilised to produce analogous formulae
for Lie superalgebras in the general linear case \cite{GIW1,GIW2,GIW3,GIW4,GIW5} 
and to a lesser extent, the orthosymplectic case \cite{GI2015}. A natural
progression of this work would be to advance the case of quantum
superalgebras. Unfortunately, for the case of quantum superalgebras, the partitioning problem
remains manifest, but no such analogous symmetry relations exist (at least in
such a straightforward manner) due to vanishing $q$-superdimensions.

In this technical article, we return to the case of quantum groups (i.e.
without underlying $\mathbb{Z}_2$-grading), specifically
$U_q[gl(n,\mathbb{C})],$ and make use of two distinct (but related) $R$-matrices,
both of which lead to an appropriate block structure of each of the
characteristic matrices, $A$ and a second, alternative characteristic matrix
$\tilde{A}$ in place of the adjoint one. As such, we provide an
alternative and {\em direct} derivation of the various formulae. Having
established such a direct construction for the non-graded quantum case, we
expect to garner insight into the derivation of formulae for the
$\mathbb{Z}_2$-graded case.

The paper is organised as follows. In Section \ref{Section2} we review certain
structural aspects of quasi-triangular Hopf algebras, especially those aspects
of the structure associated to the universal $R$-matrix. After a brief
discussion on abstract tensor operators in this context, we make use of the
preceding investigation surrounding the $R$-matrix to then apply the results to
$U_q[gl(n,\mathbb{C})]$ in Section \ref{Section3}, and to develop the pertinent
characteristic identities. In Section \ref{Section4}, the main construction is
presented, where we provide an alternative characteristic matrix $\tilde{A}$ to the adjoint matrix, 
which partitions appropriately and allows us to then derive formulae for
a variety of invariants (Section \ref{Section5}) and matrix elements (Section
\ref{Section6}). We also include five Appendices which help provide details of
our calculations for the interested reader.


\section{Quasi-triangular Hopf algebras} \label{Section2}
 
Let $H$ be a quasi-triangular Hopf algebra with universal $R$-matrix 
$$
R=\sum_ia_i\otimes b_i\equiv a_i\otimes b_i.
$$
Let $\Delta:H\rightarrow H\otimes H$ be the co-product on $H$,
$\varepsilon:H\rightarrow\mathbb{C}$ the co-unit, and $\Delta^T
= T\circ \Delta$ the opposite co-product, where $T:H\otimes H\rightarrow
H\otimes H$ is the usual twist map.
By definition $R$ satisfies
$$
R\Delta(a) = \Delta^T(a)R,\ \ \forall a\in H,
$$
and
\begin{align}
	(\Delta\otimes\id)R &= R_{13}R_{23}, \label{p1stara} \\
	(\id\otimes\Delta)R &= R_{13}R_{12}. \label{p1starb}
\end{align}
Let $\iota:\mathbb{C}\rightarrow H$ denote the unit map (so that $\iota(1)=I$, the
identity element on $H$). Applying $(\iota\circ\varepsilon\otimes\id\otimes\id)$ to (\ref{p1stara}), then
applying $m\otimes \id$ and making use of
the fundamental properties of the Hopf structure, we deduce
$$
(\iota\circ\varepsilon\otimes\id)R=I\otimes I.
$$
In a similar way, using (\ref{p1starb}), we also have
$(\id\otimes \iota\circ\varepsilon)R=I\otimes I$. Similarly applying $(S\otimes\id\otimes\id)$
or $(\id\otimes S\otimes \id)$ to (\ref{p1stara}) and then applying
$(m\otimes\id)$
gives
$$
I\otimes I = (S\otimes\id)R\cdot R = R\cdot (S\otimes \id)R,
$$
which shows explicitly that $R$ must be invertible with
\begin{align}
	R^{-1} = (S\otimes\id)R. \label{p1starstar}
\end{align}

Note that $H$ is also quasi-triangular under the opposite structure with
co-product $\Delta^T$, antipode $S^{-1}$ and $R$-matrix 
$$
R^T = T(R) = b_i\otimes a_i.
$$
Thus (\ref{p1starstar}) implies 
$$
(R^T)^{-1} = (S^{-1}\otimes\id)R^T
$$
or
\begin{align}
	R^{-1} = (\id\otimes S^{-1})R, \label{p1starstarprime}
\end{align}
so that
$$
(\id\otimes S)R^{-1} = R.
$$
Thus from (\ref{p1starstar}) we arrive at
\begin{align}
	(S\otimes S)R = (\id\otimes S)R^{-1} = R. \label{p1starstarstar}
\end{align}


\subsection{The $u$-operator}

Following Drinfeld \cite{D90} and Reshetikhin \cite{R1990}, we define 
$$
u=\sum_iS(b_i)a_i\equiv S(b_i)a_i.
$$
Then in Sweedler's \cite{S69} notation,
\begin{align}
	S(a_{(2)})u a_{(1)} = \varepsilon(a)u,
	\label{p2star}
\end{align}
with $\Delta(a) = a_{(1)}\otimes a_{(2)}$. Alternatively, we may express this
result as
$$
m(S\otimes \id)\left((I\otimes u)\Delta^T(a)\right) = \varepsilon(a)u,\ \ \forall a\in H.
$$
The result is true since
\begin{align*}
	R\Delta(a) = \Delta^T(a)R \ \Rightarrow & \ \ \ a_ia_{(1)}\otimes
	b_ia_{(2)} = a_{(2)}a_i\otimes a_{(1)}b_i\\
	\stackrel{T}{\Rightarrow} & \ \ \ b_ia_{(2)}\otimes a_ia_{(1)} =
	a_{(1)}b_i\otimes a_{(2)}a_i\\
	\stackrel{S\otimes\id}{\Rightarrow} & \ \ \ S(a_{(2)})S(b_i)\otimes a_ia_{(1)} =
	S(b_i)S(a_{(1)})\otimes a_{(2)}a_i\\
	\stackrel{m}{\Rightarrow} & \ \ \ S(a_{(2)})ua_{(1)} =
	\varepsilon(a)u.
\end{align*}
Moreover, we have the following result.
\begin{lemma}
\label{lemma1p2}
Set $R^{-1}=c_i\otimes d_i$. Then
\begin{itemize}
\item[(i)]
$S^2(a)u = ua,\ \ \forall a\in H,$
\item[(ii)]
$u$ is invertible with inverse $u^{-1}=S^{-1}(d_i)c_i.$
\end{itemize}
\end{lemma}
Proof: (i) First note that the identity 
$$
(S\otimes S)\Delta^T(a)=\Delta(S(a))
$$ 
in the Sweedler notation is 
$$
S(a_{(2)})\otimes S(a_{(1)}) =S(a)_{(1)}\otimes S(a)_{(2)}. 
$$
It follows, using (\ref{p2star}) with $a$ replaced by $S(a)$, that 
\begin{align*}
S^2(a)u &= S^2(a_{(1)})uS(a_{(2)})a_{(3)}\\
&= S(S(a)_{(2)})uS(a)_{(1)}a_{(3)}\\
&= \varepsilon(S(a)_{(1)})ua_{(2)} = ua.
\end{align*}
(ii) Explicitly 
\begin{align*}
u\cdot u^{-1} &= u\cdot S^{-1}(d_j)c_j\\
&\stackrel{(i)}{=} S(d_j)uc_j\\
&= S^{-1}(d_j)S(b_i)a_ic_j\\
&= S(b_id_j)a_ic_j\\
&= m\circ (S\otimes\mbox{id})[b_id_j\otimes a_ic_j]\\
&= m\circ (S\otimes\mbox{id})\left[ R^T\cdot(R^{-1})^T \right] = I\otimes I.
\end{align*}
\begin{flushright}$\Box$\end{flushright}

Thus Lemma \ref{lemma1p2} implies
$$
S^2(a)=uau^{-1},\ \ \forall a\in H.
$$
That is, the automorphism $S^2$ is an inner automorphism.

The canonical element $u$ has an interesting connection with $R^TR$. Before investigating this
connection, we present a technical lemma as follows.

\begin{lemma}
\label{lemma2p3}
Define two maps $\varphi,\overline{\varphi}:H^{\otimes 4}\rightarrow  H^{\otimes 2}$ by
\begin{align*}
\varphi(c_1\otimes c_2\otimes c_3\otimes c_4) &= S(c_3)c_1\otimes S(c_4)c_2,\\
\overline{\varphi}(c_1\otimes c_2\otimes c_3\otimes c_4) &= S^{-1}(c_3)c_1\otimes S(c_4)c_2.
\end{align*}
Then we have
\begin{itemize}
\item[(i)] $\varphi\left( R_{12}^TR_{23}c \right) = \varphi(c),$
\item[(ii)] $\varphi(R_{13}c) = (u\otimes I)\overline{\varphi}(c),$
\end{itemize}
$\forall c\in H^{\otimes 4}.$
\end{lemma}
Proof:
(i) In obvious notation we have
\begin{align*}
\varphi\left(R_{12}^TR_{23}c\right) 
&= \varphi(\ (b_i\otimes a_i\otimes I\otimes I)(I\otimes a_j\otimes b_j\otimes I)(c_1\otimes
c_2\otimes c_3\otimes c_4)\ ) \\
&=\varphi(b_ic_1\otimes a_ia_jc_2\otimes b_jc_3\otimes c_4)\\
&= S(b_jc_3)b_ic_1\otimes S(c_4)a_ia_jc_2\\
&= (S(c_3)\otimes S(c_4))(S(b_j)b_i\otimes a_ia_j)(c_1\otimes c_2)\\
	&= S(c_3)c_1\otimes S(c_4)c_2 = \varphi(c),
\end{align*}
since 
$$
S(b_j)b_i\otimes a_ia_j = (S\otimes \id)(S^{-1}(b_i)b_j\otimes a_ia_j)=(S\otimes \id)
\left( (R^T)^{-1} R^T\right) = I\otimes I.
$$
(ii) 
\begin{align*}
\varphi(R_{13}c) &= \varphi(\ (a_i\otimes I\otimes b_i\otimes I)(c_1\otimes c_2\otimes c_3\otimes
c_4)\ )\\
&= \varphi(a_ic_1\otimes c_2\otimes b_ic_3\otimes c_4) \\
&= S(b_ic_3)a_ic_1\otimes S(c_4)c_2\\
&= S(c_3)S(b_i)a_ic_1\otimes S(c_4)c_2\\
&= S(c_3)uc_1\otimes S(c_4)c_2\\
&= (u\otimes I) (S^{-1}(c_3)c_1\otimes S(c_4)c_2) \\
&= (u\otimes I)\ \overline{\varphi}(c).
\end{align*}
\begin{flushright}$\Box$\end{flushright}

Now we are in a position to prove the following.
\begin{thm}
$$
\Delta(u) = (u\otimes u)(R^TR)^{-1}. 
$$
\end{thm}
Proof:
\begin{align*}
\Delta(u)\cdot R^TR &= \left[ (S\otimes S)\Delta^T(b_i) \right]\cdot \Delta(a_i)\cdot R^TR\\
&= \left[ (S\otimes S)(\Delta^T(b_i)\right]\cdot R^TR\cdot \Delta(a_i)\\
&= \varphi\left( R_{12}^TR_{12}(\Delta\otimes \Delta^T)R \right),
\end{align*}
with $\varphi$ as defined in Lemma \ref{lemma2p3}. Using (\ref{p1stara}) and (\ref{p1starb}), we
have
\begin{align*}
(\Delta\otimes \Delta^T)R &= R_{13}R_{14}R_{23}R_{24}\\ 
&= R_{13}R_{23}R_{14}R_{24}\\
\Rightarrow \ \ \Delta(u)R^TR &= \varphi\left( R_{12}^TR_{12}R_{13}R_{23}R_{14}R_{24} \right)\\
&= \varphi\left( R_{12}^TR_{23}R_{13}R_{12}R_{14}R_{24} \right),
\end{align*}
where in the last step we have made use of the quantum Yang-Baxter equation
$$
R_{12}R_{13}R_{23}=R_{23}R_{13}R_{12}.
$$ 
By repeated application of Lemma \ref{lemma2p3}, we have
\begin{align*}
\Delta(u) R^TR &= \varphi(R_{13}R_{12}R_{14}R_{24})\\
&= (u\otimes I)\overline{\varphi}(R_{12}R_{14}R_{24}) \\
&= (u\otimes I)(a_ia_j\otimes S(b_k)S(b_j)b_ia_k)\\
&= (u\otimes I)(I\otimes S(b_k))(a_ia_j\otimes S(b_j)b_i)(I\otimes a_k)\\
&= (u\otimes I)(I\otimes S(b_k))[(\id\otimes S)(a_ia_j\otimes S^{-1}(b_i)b_j)](1\otimes a_k)\\
&= (u\otimes I)(I\otimes S(b_k))[(\id\otimes S)(R^{-1}\cdot R)](1\otimes a_k)\\
&= (u\otimes I)(I\otimes S(b_k)a_k)\\
&= (u\otimes I)(I\otimes u) \\
&= u\otimes u.
\end{align*}
\begin{flushright}$\Box$\end{flushright}

Since 
$$
R^{-1} = (S\otimes \id)R = S(a_i)\otimes b_i = c_i\otimes d_i,
$$
we may also write
$$
u^{-1} = S^{-1}(d_i)c_i=S^{-1}(b_i)S(a_i) = S^{-2}(b_i)a_i
$$
since $R=(S\otimes S)R$. We also observe that
$$
\tilde{R} = (R^T)^{-1}
$$
is a universal $R$-matrix for $H$. Below we shall utilise the corresponding $u$ operator, denoted
$\tilde{u}$. It is related to the $u$ operator above by the following.

\begin{lemma}
\label{lemma3p5}
$\tilde{u}=S(u^{-1}).$
\end{lemma}
Proof: 
Set 
\begin{align*}
\tilde{R} = \tilde{a}_i\otimes \tilde{b}_i &= (R^{-1})^T\\
&= [(S\otimes \id)R]^T\\
&= (\id\otimes S)R^T\\
&= b_i\otimes S(a_i). 
\end{align*}
Therefore, for the $u$-operator, we obtain
\begin{align*}
\tilde{u} = S(\tilde{b}_i)\tilde{a}_i &= S^2(a_i)b_i\\
&= S[S^{-1}(b_i)S(a_i)]\\
&= S[S^{-2}(b_i)a_i]\\
&= S(u^{-1}).
\end{align*}
\begin{flushright}$\Box$\end{flushright}

\begin{cor}
The element $S(u)u$ is central.
\end{cor}
Proof:
This follows from 
$$
S^2(a) = uau^{-1} = \tilde{u}a\tilde{u}^{-1}
$$
$\Rightarrow$ $\tilde{u}^{-1}u = S(u)u$ is central.
\begin{flushright}$\Box$\end{flushright}

Thus we also have
\begin{align*}
\Delta(\tilde{u}) &= (\tilde{u}\otimes \tilde{u}) (\tilde{R}^T\tilde{R})^{-1}\\
&= (\tilde{u}\otimes \tilde{u}) \left( R^{-1}(R^T)^{-1} \right)^{-1} \\
&=  (\tilde{u}\otimes \tilde{u}) (R^TR).
\end{align*}


\subsection{Tensor operators}

Let $V$ be a finite dimensional irreducible $H$-module and $\pi$ the representation afforded by
$V$. Given a basis $\displaystyle{\{e_\alpha\}_{\alpha=1}^n}$, $n=$dim$V$, for $V$ we define an
irreducible tensor operator of type $\pi$ to be a collection
of operators 
$$
T\equiv\left\{ T_\alpha \right\}_{\alpha=1}^n
$$
transforming according to the rule
\begin{align}
\mbox{Ad}a\circ T_{\alpha} =\pi(a)_{\beta\alpha}T_\beta,\mbox{ where } \mbox{Ad}a\circ
T_\alpha\equiv a_{(1)}T_\alpha S(a_{(2)}).
\label{starp6} 
\end{align}

\begin{lemma}
\label{lemma4p6}
$\displaystyle{T\equiv\left\{ T_\alpha \right\}_{\alpha=1}^n}$ is an irreducible tensor operator of
type $\pi$ iff
\begin{align}
aT_\alpha = T_\beta \pi(a_{(1)})_{\beta \alpha} a_{(2)}.
\label{doublestarp6}
\end{align}
\end{lemma}
Proof: If $T$ is an irreducible tensor operator then 
\begin{align*}
aT_{\alpha} &= a_{(1)}T_{\alpha}S(a_{(2)})a_{(3)}\\
&\stackrel{(\ref{starp6})}{=} \pi(a_{(1)})_{\beta\alpha}T_{\beta} a_{(2)}
\end{align*} 
which is (\ref{doublestarp6}). Conversely, if (\ref{doublestarp6}) holds then
\begin{align*}
\mbox{Ad}a\circ T_{\alpha} &=a_{(1)}T_\alpha S(a_{(2)}) \\
&\stackrel{(\ref{doublestarp6})}{=}T_{\beta}\pi(a_{(1)})_{\beta\alpha}a_{(2)} S(a_{(3)}\\
&= T_{\beta}\pi(a)_{\beta\alpha}
\end{align*}
which is (\ref{starp6}).
\begin{flushright}$\Box$\end{flushright}

Now consider the tensor operator $T$ acting on a finite dimensional module $W$. Then $T$ determines
a linear map
$$
T:V\otimes W\longrightarrow X
$$
defined by 
\begin{align}
T(e_{\alpha}\otimes w)=T_{\alpha}w,\ \ \forall w\in W,
\label{starp7}
\end{align}
where $X$ is a finite dimensional $H$-module determined by the action of $T$.

\begin{lemma}
\label{lemma5p7}
$T$ defined by (\ref{starp7}) determines an intertwining operator (or $H$-module homomorphism), i.e.
$$
T\Delta(a) = aT.
$$
\end{lemma}
Proof:
\begin{align*}
T\Delta(a)(e_{\alpha}\otimes w) &= T(a_{(1)}e_{\alpha}\otimes a_{(2)}w)\\
&= \pi(a_{(1)})_{\beta\alpha}T(e_{\beta}\otimes a_{(2)}w)\\
&\stackrel{(\ref{starp7})}{=} \pi(a_{(1)})_{\beta \alpha}T_{\beta}a_{(2)}w\\
&= aT_{\alpha}w\\
&= aT(e_\alpha\otimes w).
\end{align*}
\begin{flushright}$\Box$\end{flushright}

This shows that a tensor operator $T$ acting on $W$ is equivalent to an intertwining operator on
$V\otimes W$.

\noindent
\underline{Example}: $R^TR$ commutes with the action of $\Delta$. Hence we have a tensor operator
defined by 
$$
T_{\alpha\beta}=\pi(b_ia_j)_{\alpha\beta} a_ib_j.
$$ 
\begin{flushright}$\Box$\end{flushright}

To understand the nature of this tensor operator, of importance below, we need to consider some
module constructions. 

First, given finite dimensional $H$-modules $V$, $W$, the space of linear maps
from $V$ to $W$, denoted $\ell(V,W)$, is also an $H$-module under the
action defined by 
\begin{align}
(a\circ f)(v) = a_{(1)}f(S(a_{(2)})v),\ \  v\in V, f\in \ell(V,W).
\label{starp8}
\end{align}
Note that $f$ is an $H$-module homomorphism iff it is an invariant under this action, i.e., 
$$
a_{(1)}f(S(a_{(2)})v) = \varepsilon(a) f(v).
$$
In the case $W=\mathbb{C}$ is the trivial 1-dimensional module, the action (\ref{starp8}) reduces to
$$
(a\circ f)(v) = \varepsilon(a_{(2)})f(S(a_{(1)})v), 
$$
i.e. 
$$
(a\circ f)(v) = f(S(a)v),\ \ \forall f\in V^*.
$$
This shows that a dual module is also a module under the action determined by the antipode.

\begin{prop}
\label{page8prop1}
Let $V$, $W$ be finite-dimensional $H$-modules. Then we have an $H$-module isomorphism
$$
W\otimes V^* \cong \ell(V,W).
$$
\end{prop}
Proof: Following the standard argument, we define a linear map $\varphi:W\otimes V^*\longrightarrow
\ell(V,W)$ by
$$
\varphi(w\otimes v^*)(u) = v^*(u)w, \ \ \forall u\in V,\ w\in W,\ v^*\in V^*.
$$
Then $\varphi$ is well-defined and one-to-one. It is also onto since given any $f\in\ell(V,W)$ we
may write (sum on $\alpha$)
$$
f= \varphi(f(e_\alpha)\otimes e_\alpha^*)
$$
where $\{ e_\alpha \}$ is a basis for $V$ with dual basis $\{ e_\alpha^*\}$ for $V^*$ defined by
$e_\alpha^*(e_\beta) = \delta_{\alpha \beta}$. Indeed,
$$
\varphi(f(e_\alpha)\otimes e_\alpha^*)(u) = e_\alpha^*(u)f(e_\alpha) = f(u),\ \ \forall u\in V.
$$
It remains to show that $\varphi$ is an $H$-module homomorphism. To this end we have
\begin{align*}
\varphi(a\circ(w\otimes v^*))(u) &= \varphi(a_{(1)}w\otimes a_{(2)}v^*)(u)\\
&= (a_{(2)}v^*)(u)a_{(1)}w\\
&= v^*(S(a_{(2)}u))a_{(1)}w\\
&= a_{(1)}\varphi(w\otimes v^*)(S(a_{(2)}u)\\
&= a\circ(\varphi(w\otimes v^*))(u).
\end{align*}
\begin{flushright}$\Box$\end{flushright}

In the case that $W=V$, we arrive at the following.
\begin{prop} \label{Prop2}
If $V$ is a finite dimensional irreducible $H$-module the the identity module occurs exactly once in
$V\otimes V^*$ and is spanned by the vector (notation as above)
$$
\xi = e_\alpha\otimes e_\alpha^*,
$$
with summation on the label $\alpha$.
\end{prop}
Proof: Since $V\otimes V^*\cong \ell(V,V)$, Schur's lemma implies the identity module occurs exactly once in
$V\otimes V^*$ (all invariants in $\ell(V,V)$ are scalar multiples of the identity by classical
arguments). Then
\begin{align*}
a\circ \xi &= a_{(1)}e_\alpha\otimes a_{(2)}e_\alpha^*\\
&= a_{(1)}e_\alpha\otimes \langle a_{(2)}e_\alpha^*,e_\beta \rangle e_\beta^*\\
&= \langle e_\alpha^*,S(a_{(2)})e_\beta \rangle a_{(1)}e_\alpha\otimes e_\beta^*\\
&= a_{(1)}S(a_{(2)})e_\beta\otimes e_\beta^*\\
&= \varepsilon(a)\xi. 
\end{align*} 
\begin{flushright}$\Box$\end{flushright}

\indent
\underline{Note}: 
The uniqueness of the identity representation in $V\otimes V^*$ also applies to
$V^*\otimes V$ (see Appendix B for details). In fact, using the $R$-matrix it is straightforward
to show that the two spaces $V\otimes V^*$ and $V^*\otimes V$ are isomorphic. 


\subsection{Tensor operators from $R^TR$}

Let $\pi$ be the representation afforded by a finite dimensional irreducible $H$-module $V$. Then
$T\equiv (\pi\otimes \mbox{id})R^TR$ determines a tensor operator with components
$$
T_{\alpha\beta} = (\zeta_{\alpha\beta}\otimes \mbox{id})R^TR,
$$
where $\zeta_{\alpha\beta}:H\rightarrow \mathbb{C}$ is such that
$\zeta_{\alpha\beta}(a)=\pi(a)_{\alpha\beta}$. Given another $H$-module $W$, the above components
are defined by
$$
(\pi\otimes\mbox{id})R^TR(e_\beta\otimes w) = e_\alpha\otimes T_{\alpha\beta}w
$$
so
$$
T_{\alpha\beta}w = (e_\alpha^*\otimes \mbox{id})R^TR(e_\beta\otimes w).
$$
\begin{prop}
The map $\varphi:V^*\otimes V\otimes W\longrightarrow W$ defined by
$$
\varphi(e_\alpha^*\otimes e_\beta\otimes w) = T_{\alpha\beta}w
$$
is an intertwining operator ($H$-module homomorphism).
\end{prop}
Proof: We have, for any $a\in H$,
\begin{align*}
\varphi\cdot a (e_\alpha^*\otimes e_\beta\otimes w) &= \varphi(a_{(1)}e_\alpha^*\otimes
a_{(2)}e_\beta\otimes a_{(3)}w)\\
&= (a_{(1)}e_\alpha^*\otimes\mbox{id})R^TR(a_{(2)}e_\beta\otimes a_{(3)}w)\\
&=(a_{(1)}e_\alpha^*\otimes\mbox{id})R^TR\Delta(a_{(2)})(e_\beta\otimes w)\\
&= (e_\alpha^*\otimes \mbox{id})\cdot (S(a_{(1)})\otimes I)\Delta(a_{(2)})R^TR(e_\beta\otimes w)  \\
&= (e_\alpha^*\otimes \mbox{id})\cdot(S(a_{(1)})a_{(2)}\otimes a_{(3)})R^TR(e_\beta\otimes w)  \\
&=  (e_\alpha^*\otimes \mbox{id})\cdot(I\otimes a)R^TR(e_\beta\otimes w)  \\
&= aT_{\alpha\beta}w\\
&= a\cdot\varphi(e_\alpha^*\otimes w).
\end{align*}
\begin{flushright}$\Box$\end{flushright}

In terms of components we have 
\begin{align*}
aT_{\alpha\beta}w &= \varphi(a_{(1)}e_\alpha^*\otimes a_{(2)}e_\beta\otimes a_{(3)}w)\\
&= \pi^*(a_{(1)})_{\gamma\alpha}\pi(a_{(2)})_{\delta\beta}\varphi(e_\gamma^*\otimes e_\delta\otimes
a_{(3)}w)\\
&= \pi^*\otimes \pi (\Delta(a_{(1)})_{\gamma\alpha,\delta\beta}T_{\gamma\delta}a_{(2)}w
\end{align*}
or
\begin{align}
aT_{\alpha\beta} &= \pi^*\otimes
\pi(\Delta(a_{(1)}))_{\gamma\alpha,\delta\beta}T_{\gamma\delta}a_{(2)}.
\label{p11star}
\end{align}
This shows explicitly that the entries of the matrix $(\pi\otimes\id)R^TR$ transform as a tensor
operator of type $\pi^*\otimes \pi$.

\noindent
\underline{Notes}: (1) If $\{e_\alpha\}$ is a basis for $V$, assumed above, then
\begin{align*}
ae_\alpha &= \pi(a)_{\beta\alpha}e_\beta\\
&= \langle e_\beta^*,ae_\alpha\rangle e_\beta\\
\Rightarrow \ \ \pi(a)_{\beta\alpha} &= \langle e_\beta^*,ae_\alpha\rangle,
\end{align*}
for $a\in H$.\\
(2) If $\{e_\alpha^*\}$ is the corresponding dual basis for $V^*$, as above, then we have
$$
ae_\alpha^* = \pi^*(a)_{\beta\alpha}e_\beta^*.
$$
On the other hand,
\begin{align*}
ae_\alpha^* &= \langle e_\beta,ae_\alpha^* \rangle e_\beta^*\\
&= \langle S(a)e_\beta,e_\alpha^* \rangle e_\beta^*\\
&= \pi(S(a))_{\alpha\beta} e_\beta^*\\
	\Rightarrow \ \ \pi^*(a) &= \pi^t(S(a)),\ \ \forall a\in H, 
\end{align*}
which is the usual definition of dual representation adopted below.


\section{Quantum groups} \label{Section3}

A well-known example of a quasi-triangular Hopf algebra is the quantum group $U_q(L)$ associated
with a simple Lie algebra $L$ with co-unit and co-product given by 
$$
\varepsilon(x)=0,\ \ x=e_i,\ f_i,\ h_i, 
$$
$$
\Delta (x) = q^{h_i/2}\otimes x + x\otimes q^{-h_i/2},\ \ x=e_i,\ f_i,\ \ \Delta(h_i)=h_i\otimes I +
I\otimes h_i,
$$
and antipode
$$
S(a) = q^{-h_\rho}\gamma(a)q^{h_\rho},\ \ a\in U_q(L)
$$
where $\gamma$ is the principal anti-homomorphism defined by
$$
\gamma(I)=I,\ \ \gamma(x)=-x,\ \ x=e_i,\ f_i,\ h_i.
$$
In this case the $R$-matrix is expressible
\begin{align}
R &= q^{h_i\otimes h^i}\left\{ I + \sum_s e_s\otimes e^s \right\} \ \in U_q^{(-)}(L)\otimes
U_q^{(+)}(L), 
\label{p12star}
\end{align}
with $\{h_i| i=1,\ldots,\ell \}$ a basis for the Cartan subalgebra of $L$ with dual basis
$\{h^i|i=1,\ldots,\ell\}$ under a given invariant bi-linear form $(\ ,\ )$ unique up to scalar
multiples on $L$. Note here that $\rho$ is half the sum of positive roots of $L$, and $h_\rho$ is
defined with respect to $(\ ,\ )$, for all $\mu$ in the dual space of the Cartan
subalgebra, as
$$
\mu(h_\rho) = (\mu,\rho).
$$
Algebraic relations satisfied by the generators are given, for example, in
\cite{ZGB91}.

Observe that
$$
S^2(a) = q^{-2h_\rho} a q^{2h_\rho},\ \ a\in U_q(L)
$$
which implies that $v=q^{2h_\rho} u$
is a central element with inverse (also central)
$$
v^{-1} = u^{-1}q^{-2h_\rho}.
$$
Moreover, since $q^{\pm 2h_\rho}$ are group-like, we also have
$$
\Delta(v) = (v\otimes v)(R^TR)^{-1}
$$
or
$$
R^TR = (v\otimes v)\Delta(v^{-1}).
$$
We call $U_q(L)$ a {\em Ribbon} Hopf algebra.

In terms of the $R$-matrix 
$$
\tilde{R} = (R^T)^{-1}
$$
we have the corresponding $u$-operator
$$
\tilde{u} = S(u^{-1})
$$
and central element
\begin{align*}
\tilde{v} &= q^{2h_\rho}\tilde{u}\\
&= q^{2h_\rho} S(u^{-1}) \\
&= S(u^{-1}q^{-2h_\rho})\\
&= S( (q^{2h_\rho}u)^{-1})\\
&= S(v^{-1}).
\end{align*}
In this case
\begin{align*}
\Delta(\tilde{v}) &= (\tilde{v}\otimes \tilde{v}) (\tilde{R}^T\tilde{R})^{-1}\\
&= (\tilde{v}\otimes \tilde{v}) (R^{-1} (R^T)^{-1})^{-1}\\
&= (\tilde{v}\otimes \tilde{v}) R^TR\\
&= S(v^{-1})\otimes S(v^{-1}) R^TR.
\end{align*}


\subsection{Eigenvalues of $v$}

Let $V(\Lambda)$ be a finite dimensional irreducible module with highest weight $\Lambda\in D_+$,
and let $e_-^\Lambda$ be the minimal weight vector. Then observe that with $R$ as in
(\ref{p12star}), 
\begin{align*}
ue_-^\Lambda &= q^{-h_ih^i}e_-^\Lambda = q^{-(\Lambda_-,\Lambda_-)}e_-^\Lambda \\
\Rightarrow \ \ ve_-^\Lambda &= q^{2h_\rho}ue_-^\Lambda =
q^{-(\Lambda_-,\Lambda_-)+(\Lambda_-,2\rho)}e_-^\Lambda.
\end{align*}
Now observe that there is a unique Weyl group element $\tau$ sending positive roots to negative
roots so that
$$
\Lambda_- = \tau(\Lambda).
$$
It follows that 
\begin{align*}
(\Lambda_-,2\rho) - (\Lambda_-,\Lambda_-) &= (\Lambda,\tau(2\rho)) - (\tau(\Lambda),\tau(\Lambda)) \\
&= -(\Lambda,2\rho) - (\Lambda,\Lambda)\\
&= -(\Lambda,\Lambda+2\rho).
\end{align*}
Consequently, on $V(\Lambda)$ the invariant $v$ takes the constant value
\begin{align}
\chi_\Lambda(v) &= q^{-(\Lambda,\Lambda+2\rho)}.
\label{p14star}
\end{align}
On the other hand, $\tilde{v}=S(v^{-1})$ takes the eigenvalue
\begin{align}
\chi_\Lambda(\tilde{v}) &= q^{(\Lambda,\Lambda+2\rho)}
\label{p14stardash}
\end{align}
as may be verified directly by action on the highest weight vector.

\noindent
\underline{Note}: It follows that 
$$
\frac{1-q^{-(\Lambda,\Lambda+2\rho)}}{q-q^{-1}} \ \longrightarrow \ \frac12(\Lambda,\Lambda+2\rho)
$$
in the limit $q\rightarrow 1$, so we may take
$$
C_q = 2\left( \frac{1-v}{q-q^{-1}} \right)
$$
as a generalised quadratic invariant. 

The properties of such invariants and their relationship to those considered below is worth further
consideration. Note that
$$
\Delta(C_q) = \frac{2}{q-q^{-1}}\left( I\otimes I - \Delta(v) \right) = \frac{2}{q-q^{-1}}\left(
I\otimes I - (v\otimes v)(R^TR)^{-1} \right).
$$


\subsection{Characteristic identities}

Let $V(\Lambda)$ be a finite dimensional irreducible $U_q(L)$-module with highest weight $\Lambda\in
D_+$ and $\pi_\Lambda$ the representation afforded by $V(\Lambda)$.Then we have the tensor matrix
$$
A = (q-q^{-1})^{-1}\pi_\lambda\otimes \id(I\otimes I - R^TR)
$$
which reduces to the characteristic matrix for the semisimple Lie algebra $L$ previously considered
by Green and Bracken \cite{Green1971,BraGre1971} and Gould \cite{Gould1978,Gould1980}, in the limit $q\rightarrow 1$. We have seen that the entries
$A_{\alpha\beta}$ of the matrix $A$ transform as a tensor operator of type $\pi_\Lambda^*\otimes
\pi_\Lambda$.

On the other hand, acting on a finite dimensional module $V(\mu)$, $A$ may be regarded as an
intertwining operator on the tensor product module $V(\Lambda)\otimes V(\mu)$:
\begin{align*}
A &= (q-q^{-1})^{-1}\pi_\lambda\otimes \pi_\mu(I\otimes I - R^TR)\\
 &= (q-q^{-1})^{-1}\pi_\lambda\otimes \pi_\mu(I\otimes I -(v\otimes v)\Delta(v^{-1}) ).
\end{align*}
The module $V(\Lambda)\otimes V(\mu)$ is well known to be completely reducible into a direct sum of
irreducible modules with highest weights $\mu+\lambda_i$, with $\lambda_i$ a weight occuring in
$V(\Lambda)$. It follows that if $\lambda_1,\ldots,\lambda_k$ are the {\em distinct} weights in
$V(\Lambda)$ then $A$ satisfies the polynomial identity 
\begin{align}
\prod_{i=1}^k\left[ A-a_i \right] = 0,
\label{p15star}
\end{align}
where 
\begin{align*}
a_i &=
(q-q^{-1})^{-1}(1-q^{(\mu+\lambda_i,\mu+\lambda_i+2\rho)-(\Lambda,\Lambda+2\rho)-(\mu,\mu+2\rho)})\\
&= \frac{1-q^{2\alpha_i}}{q-q^{-1}},
\end{align*}
where
\begin{align*}
\alpha_i &= \frac12\left[
(\mu+\lambda_i,\mu+\lambda_i+2\rho)-(\Lambda,\Lambda+2\rho)-(\mu,\mu+2\rho) \right]\\
&= \frac12\left[ (\lambda_i,\lambda_i + 2(\mu+\rho) - (\Lambda,\Lambda+2\rho) \right]
\end{align*}
are the classical characteristic roots. These are a quantum analogue of the characteristic
identities considered by Green and Bracken \cite{Green1971,BraGre1971}, Gould
\cite{Gould1978,Gould1980} and others \cite{OBCC77} for a simple Lie algebra $L$. 


\subsection{Characteristic identities for $U_q[gl(n,\mathbb{C})]$}

Recall that $U_q(n)=U_q[gl(n,\mathbb{C})]$ is the quantum group with simple generators
$$
e_i = E_{i\ i+1},\ f_i=E_{i+1\ i},\ h_i=E_{ii} - E_{i+1\ i+1},\ 1\leq i\leq n.
$$
Here we consider the vector module $V_0$ of $gl(n,\mathbb{C})$ and $U_q(n)$ which
is undeformed and irreducible of highest weight $\varepsilon_1$. The corresponding representation is
defined by 
$$
\pi_0(E_{i\ i+1}) = e_{i\ i+1},\ \pi_0(E_{i+1\ i})=e_{i+1\ i},\ \pi_0(E_{ii}) = e_{ii},
$$
where, as usual, $e_{ij}$ is an elementary matrix.

In this case we have, from previous work of Jimbo \cite{J85} and others
\cite{GZB90,LG92},
\begin{align}
(\pi_0\otimes\id)R &= \sum_{i\leq j} e_{ji}\otimes \hat{E}_{ij},\label{p16stara}\\
(\pi_0\otimes\id)R^T &= \sum_{i\leq j} e_{ij}\otimes \hat{E}_{ji},\label{p16starb}
\end{align} 
where
\begin{align}
\hat{E}_{ij} &= \left\{ \begin{array}{rl}  
(q-q^{-1})q^{\frac12(E_{ii}+E_{jj}-1)}E_{ij}, & i\neq j\\
q^{E_{ii}}, & i=j
\end{array}
\right.
\label{p17star}
\end{align}
with $E_{ij}$ defined recursively according to
\begin{align}
E_{ij} &= E_{ik}E_{kj} - q^{-1}E_{kj}E_{ik},\ \ i\lessgtr k \lessgtr j.
\label{p17stardash}
\end{align}
This gives the characteristic matrix
\begin{align}
\overline{A}_{ij} &= (q-q^{-1})\left\{ \delta_{ij} - (\pi_0)_{ij}\otimes \id (R^TR) \right\}
\nonumber\\
&= (q-q^{-1})\left\{ \delta_{ij} - \sum_{k\geq i\vee j}^n
	\hat{E}_{ki}\hat{E}_{jk} \right\},
\label{p17doublestar}
\end{align}
where $i\vee j=\mbox{max}(i,j)$. This characteristic matrix is referred to as the adjoint matrix in reference \cite{GLB1992}.

\noindent
\underline{Note:} The operator (\ref{p17star}) should not be confused with
\begin{align}
\tilde{E}_{ij} &= \left\{ \begin{array}{rl}  
-(q-q^{-1})q^{-\frac12(E_{ii}+E_{jj}-1)}E_{ij}', & i\neq j\\
q^{-E_{ii}}, & i=j
\end{array}
\right.
\label{p17triplestar}
\end{align}
with $E_{ij}'$ defined recursively by
$$
E_{ij}' = E_{ik}'E_{kj}' - qE_{kj}'E_{ik}'
$$
appearing in reference \cite{GLB1992}.

In this case the antipode is given by
$$
S^\pm(e_i) =-q^{\mp h_\rho}e_iq^{\pm h_\rho} = -q^{\mp 1}e_i,\ \
S^\pm(f_i)=-q^{\pm 1}f_i,\ S(E_{ii})
= -E_{ii},
$$
so the transformation properties of equation (\ref{p11star}) reduce to 
\begin{align*}
q^{E_{ii}}\overline{A}_{k\ell} &=
\overline{A}_{k\ell}q^{E_{ii}+(\varepsilon_i,\varepsilon_\ell-\varepsilon_k)},\ 1\leq k,\ell\leq n,\\
e_i\overline{A}_{k\ell} &=
	q^{\frac12(\varepsilon_i-\varepsilon_{i+1},\varepsilon_\ell-\varepsilon_k)}\overline{A}_{k\ell}e_i
	+ \left\{ \delta_{\ell \ i+1}
	q^{-\frac12(\varepsilon_i-\varepsilon_{i+1},\varepsilon_k)}\overline{A}_{ki} -
	q^{-1}q^{-\frac12(\varepsilon_i-\varepsilon_{i+1},\varepsilon_\ell)}\delta_{ki} \overline{A}_{i+1\ \ell} \right\}
q^{-h_i/2},\\
f_i \overline{A}_{k\ell} &=
	q^{\frac12(\varepsilon_i-\varepsilon_{i+1},\varepsilon_\ell-\varepsilon_k)}\overline{A}_{k\ell}f_i 
	+ \left\{ \delta_{i\ell}q^{\frac12(\varepsilon_i-\varepsilon_{i+1},\varepsilon_k)}\overline{A}_{k\ i+1} - q\delta_{k\
	i+1}q^{-\frac12(\varepsilon_i-\varepsilon_{i+1},\varepsilon_\ell)}\overline{A}_{i\ell} \right\} q^{-h_i/2}
\end{align*}
or
\begin{align}
[e_i,\overline{A}_{k\ell}]_{q_i} &=   \left\{ \delta_{\ell \ i+1}
	q^{-\frac12(\varepsilon_i-\varepsilon_{i+1},\varepsilon_k)}\overline{A}_{ki} -
	q^{-1}q^{-\frac12(\varepsilon_i-\varepsilon_{i+1},\varepsilon_\ell)}\delta_{ki} \overline{A}_{i+1\ \ell} \right\}
q^{-h_i/2}, \label{p18triplestara}\\
[f_i,\overline{A}_{k\ell}]_{q_i} &= \left\{
	\delta_{i\ell}q^{\frac12(\varepsilon_i-\varepsilon_{i+1},\varepsilon_k)}\overline{A}_{k\ i+1} - q\delta_{k\
	i+1}q^{-\frac12(\varepsilon_i-\varepsilon_{i+1},\varepsilon_\ell)}\overline{A}_{i\ell} \right\} q^{-h_i/2},
\label{p18triplestarb}
\end{align}
where the $q$-bracket on the LHS is defined by
$$
[x,\overline{A}_{k\ell}]_{q_i} = x\overline{A}_{k\ell} -
q^{\frac12(\varepsilon_i-\varepsilon_{i+1},\varepsilon_\ell-\varepsilon_k)}\overline{A}_{k\ell}x,\ \ x=e_i,f_i.
$$
We note here that $\varepsilon_i-\varepsilon_{i+1}$ denotes the $i$th simple
root.


\subsection{Polynomial identities}

In this case the identities (\ref{p15star}) reduce to
\begin{align}
\prod_{r=1}^n(\overline{A} - \overline{a}_r)=0,
\label{p18star}
\end{align}
where
$$
\overline{a}_r = \frac{1-q^{2\overline{\alpha}_r}}{q-q^{-1}},
$$
where 
$$
\overline{\alpha}_r =\frac12\left[ (\varepsilon_r,\varepsilon_r+2(\Lambda+\rho)) -
(\varepsilon_1,\varepsilon_1+2\rho) \right]. 
$$
In this case
$$
\rho = \frac12\sum_{i=1}^n(n+1-2i)\varepsilon_i
$$
so that
\begin{align*}
\overline{\alpha}_r &= \frac12\left[ 1+2\Lambda_r+(n+1-2r)-1-(n+1-2) \right] \\
&= \Lambda_r+1-r
\end{align*}
which are the usual classical adjoint roots.

The above polynomial identities allow the construction of projection operators
$$
\overline{P}_r = \prod_{\ell\neq r}\left( \frac{\overline{A}-\overline{a}_r}{\overline{a}_\ell -
\overline{a}_r} \right)
$$
whose entries are polynomials in the generators which determine squares of {\em Wigner coefficients}
(WCs). Of particular interest is the $(n,n)$ entry
\begin{align}
\overline{\omega}_r &= \left( \overline{P}_r \right)_{nn}
\label{p19star}
\end{align}
which in fact determines an invariant of $U_q(n-1)$ whose eigenvalues are squares of
{\em reduced Wigner coefficients} (RWCs). Specifically, following the notation
of \cite{Gould1992}, on an
irreducible $U_q(n-1)$-module $V(\Lambda_0)$ with highest weight $\Lambda_0$
contained in $V(\Lambda)$, the eigenvalue of the invariants (\ref{p19star}) are given by
\begin{align}
\chi_{(\Lambda,\Lambda_0)}(\overline{\omega}_r) &= 
\left| 
\left.
\left\langle 
  \begin{array}{c}
    \Lambda\\
    \Lambda_0+\varepsilon_r
  \end{array}
\right.
\right|
\left|
 \begin{array}{c}
    \varepsilon_1 \\
    0
  \end{array}
\right.
;
\left.
  \begin{array}{c}
    \Lambda\\
    \Lambda_0
  \end{array}
\right\rangle
\right|^2.
\label{p19doublestar}
\end{align}

Our aim, following the approach of references \cite{GLB1992,Gould1992}, is to utilise the above characteristic identities
(\ref{p18star}) to evaluate the squared RWCs (\ref{p19doublestar}). Unfortunately this approach does
not apply in this case since the matrix $\overline{A}$ of equation (\ref{p17star}) does not have the
correct block structure. Specifically, the first $n-1$ rows and columns do not give the
corresponding adjoint matrix $\overline{A}$ for $U_q(n-1)$ which is a necessary
requirement for the application of this approach. Here we present an alternative construction which
remedies this deficiency.


\section{An alternative construction} \label{Section4}

Here we utilise the alternative $R$-matrix $\tilde{R}=(R^T)^{-1}$. Remarkably the $\overline{A}$
matrix constructed from this $R$-matrix has the required block structure, necessary to implement the
approach of \cite{GLB1992,Gould1992}, as we will see.

Using 
\begin{align*}
\tilde{R} &= (R^T)^{-1} = [(S\otimes \id)R]^T = (\id\otimes S)R^T,\\
\tilde{R}^T &= R^{-1} = (\id\otimes S^{-1})R,
\end{align*}
we have, making use of (\ref{p16stara}),(\ref{p16starb}), the $L$-operators 
\begin{align*}
(\pi_0\otimes \id)\tilde{R} &= (\pi_0\otimes\id)(\id\otimes S)R^T = \sum_{i\leq j}e_{ij}\otimes
S(\hat{E}_{ji}),\\
(\pi_0\otimes\id)\tilde{R}^T &= (\pi_0\otimes\id)(\id\otimes S^{-1})R = \sum_{i\leq j}e_{ji}\otimes
S^{-1}(\hat{E}_{ij}).
\end{align*}
Therefore we arrive at the new characteristic matrix
$$
\tilde{A}_{ij} = (q-q^{-1})^{-1}\left\{ \delta_{ij} - ({\pi_0}_{ij}\otimes\id)(\tilde{R}^T\tilde{R}) \right\}
$$
where now 
\begin{align*}
(\pi_0\otimes\id)(\tilde{R}^T\tilde{R}) &= \sum_{i\leq j,\ k\leq \ell} e_{ji}e_{k\ell}\otimes
S^{-1}(\hat{E}_{ij})S(\hat{E}_{\ell k})\\
&= \sum_{k\leq i\wedge j} e_{ij}\otimes S^{-1}(\hat{E}_{ki})S(\hat{E}_{jk}).
\end{align*}
Thus we arrive at the alternative characteristic matrix
$$
\tilde{A}_{ij} = (q-q^{-1})^{-1}\left\{ \delta_{ij} - \sum_{k\leq i\wedge j}
S^{-1}(\hat{E}_{ki})S(\hat{E}_{jk})\right\},
$$
where $i\wedge j=\mbox{min}(i,j)$. Note that the transformation properties
(\ref{p18triplestara}),(\ref{p18triplestarb}) also hold for this matrix.

In this case the matrix $\tilde{A}$ may be written in invariant form as
$$
\tilde{A} = (q-q^{-1})^{-1}(\pi_0\otimes \id)\left[ I\otimes I -
(\tilde{v}\otimes\tilde{v})\Delta(\tilde{v}^{-1}) \right]
$$
so that, in view of equation (\ref{p14star}), on an irreducible module $V(\Lambda)$ the matrix
$\tilde{A}$ satisfies the polynomial identity
$$
\prod_{r=1}^n(\tilde{A} - \tilde{a}_r)=0,
$$
where now 
$$
\tilde{a}_r = (q-q^{-1})^{-1}[1-q^{-2\overline{\alpha}_r}]
$$
with $\overline{\alpha}_r=\Lambda_r+1-r$ the classical adjoint roots as before.


\subsection{Partitioning of matrix $\tilde{A}$: $U_q(n)\supset U_q(n-1)$}

We observe that the first $n-1$ rows and columns of the matrix $\tilde{A}$
yields the corresponding
$U_q(n-1)$ matrix, denoted $\tilde{A}_0$:
$$
\left( \begin{array}{ccc:c} 
&&& \\ 
\ & \tilde{A}_0 & & \tilde{A}_{in} \\ 
&&& \\ 
\hdashline
&&& \\
 & \tilde{A}_{ni} & & \tilde{A}_{nn} \end{array} \right).
$$
Now observe that the elementary generators of $U_q(n-1)$ are given by $e_i,f_i,h_i$, $1\leq
i<n-1$ so the transformation law of equations (\ref{p18triplestara}),(\ref{p18triplestarb}) in the
case $k=\ell=n$ gives
\begin{align*}
q^{E_{ii}}\tilde{A}_{nn} &= \tilde{A}_{nn}q^{E_{ii}},\ 1\leq i<n,\\
e_i\tilde{A}_{nn} - \tilde{A}_{nn}e_i &= f_i\tilde{A}_{nn}-\tilde{A}_{nn}f_i=0,\ \ 1\leq i<n-1,
\end{align*}
so that the entry $\tilde{A}_{nn}$ is a $U_q(n-1)$ invariant.

Now consider the entries
\begin{align}
\phi_i &\equiv \tilde{A}_{in},\ \ 1\leq i<n
\label{p23star}
\end{align}
of the last column of the matrix $\tilde{A}$. Setting $\ell=n$ in equations
(\ref{p18triplestara}),(\ref{p18triplestarb}) gives
\begin{align*}
q^{E_{ii}}\phi_k &= \phi_kq^{E_{ii}-\delta_{ik}} = q^{-\delta_{ik}} \phi_k q^{E_{ii}},\\
	e_i\phi_k - q^{-\frac12(\varepsilon_i-\varepsilon_{i+1},\varepsilon_k)} \phi_ke_i &=
-q^{-1}\delta_{ki}\phi_{i+1}q^{-h_i/2},\\
	f_i\phi_k - q^{-\frac12(\varepsilon_i-\varepsilon_{i+1},\varepsilon_k)}\phi_kf_i &= -q\delta_{k\ i+1}\phi_iq^{-h_i/2}.
\end{align*}
By comparison with Appendix A, we see that the operators (\ref{p23star}) transform as a dual vector
operator of $U_q(n-1)$ (i.e. a tensor operator of rank $\pi_0^*$).

Similarly for the entries of the last row
\begin{align}
\psi_i \equiv \tilde{A}_{ni},\ 1\leq i<n,
\label{p23doublestar}
\end{align}
we have, setting $k=n$ into equations (\ref{p18triplestara}),(\ref{p18triplestarb}), the
transformation law
\begin{align*}
q^{E_{ii}}\psi_\ell &= q^{\delta_{i\ell}}\psi_\ell q^{E_{ii}},\\
	e_i\psi_\ell - q^{\frac12(\varepsilon_i-\varepsilon_{i+1},\varepsilon_\ell)}\psi_\ell e_i &= \delta_{\ell\
i+1}\psi_iq^{-h_i/2},\\
	f_i\psi_\ell - q^{\frac12(\varepsilon_i-\varepsilon_{i+1},\varepsilon_\ell)}\psi_\ell f_i &= \delta_{\ell i} \psi_{i+1}
q^{-h_i/2}.
\end{align*}
Again, by comparison with Appendix A, it follows that the operators (\ref{p23doublestar}) transform as a
vector operator of $U_q(n-1)$.


\subsection{$U_q(n)$ vector operators and shift components}

The $U_q(n)$ projection operators
\begin{align}
\tilde{P}_r = \prod_{\ell\neq r}^n\left( \frac{\tilde{A} - \tilde{a}_\ell}{\tilde{a}_r -
\tilde{a}_\ell} \right)
\label{p24star}
\end{align}
form an orthogonal set of idempotents adding up to the identity:
$$
\tilde{P}_r\tilde{P}_\ell = \delta_{r\ell}\tilde{P}_r,\ \ \sum_{r=1}^n \tilde{P}_r = I.
$$
As an operator on the tensor product module $V_0\otimes V(\Lambda)$, $\tilde{P}_r$ projects onto the
submodule
$$
V(\Lambda + \varepsilon_r)\subset V_0\otimes V(\Lambda).
$$
On the other hand, $\tilde{P}_r$ may be viewed as an $n\times n$ matrix with entries
$(\tilde{P}_r)_{ij}$ defined by
$$
\tilde{P}_r(e_i\otimes v) = e_j\otimes (\tilde{P}_r)_{ji}v,\ \ v\in V(\Lambda).
$$
\underline{Note:} With this definition, powers of the matrix $\tilde{A}$ are given recursively by
$$
\left( \tilde{A}^{m+1} \right)_{ij} = (\tilde{A})_{ik}\left(\tilde{A}^m\right)_{kj} =
\left(\tilde{A}^m\right)_{ik}(\tilde{A})_{kj},
$$
with summation on $k$ from 1 to $n$.

If $\psi_i$ is a vector operator on $V(\Lambda)$ then we have seen that $\psi$ on $V_0\otimes
V(\Lambda)$, defined by
$$
\psi(e_i\otimes v) = \psi_iv,\ \ \forall v\in V(\Lambda)
$$
is an intertwining operator. So too is $\psi[r]$ defined by
\begin{align*}
\psi[r][(e_i\otimes v) &= \psi\tilde{P}_r(e_i\otimes v)\\
&= \psi(e_j\otimes (\tilde{P}_r)_{ji}v)\\
&= \psi_j(\tilde{P}_r)_{ji}v.
\end{align*}
Thus 
$$
\psi[r]_i \equiv \psi_j(\tilde{P}_r)_{ji}
$$
also gives rise to a vector operator which increases the highest weight $\Lambda$ by
$\varepsilon_r,$
i.e. affects the shift $\Lambda\rightarrow\Lambda+\varepsilon_r.$ We call $\psi[r]_i$ a shift vector
operator.

By contrast, if $\phi_i$ is a dual vector operator on $V(\Lambda)$, $\tilde{P}_r$ projects out the
shift components from the {\em left} (see Appendix B):
$$
\phi[r]_i = (\tilde{P}_r)_{ij}\phi_j
$$
which affects the shifts $\Lambda \rightarrow \Lambda-\varepsilon_r$.

To project out the shift components of vector (respectively dual vector) operators from the left
(respectively right), however, we also require the $U_q(n)$ vector matrix and its characteristic
identity. These matrices are of particular importance for determining dual vector WCs and RWCs.


\subsection{The vector matrix revisited}

In what follows, $\overline{\pi}_0$ denotes the (undeformed) ``dual'' vector representation given by
\begin{align*}
& \overline{\pi}_0 (E_{i\ i+1}) = \overline{\pi}_0(e_i) = -e_{i+1\ i},\\
& \overline{\pi}_0 (f_i) = -e_{i\ i+1},\ \ \overline{\pi}_0(E_{ii}) = -e_{ii}.
\end{align*}
Like $\pi_0$, this representation is undeformed and unitary (see Appendix E) unlike the dual vector
representation $\pi_0^*$. The dual vector representation is defined by
$$
\pi_0^*(a) = \pi_0^t(S(a)),\ \ \forall a\in U_q(n),
$$
whereas 
$$
\overline{\pi}_0(a) = \pi_0^t(\gamma(a)),
$$
with $\gamma$ the principal anti-automorphism. The two representations are related by
\begin{align*}
\overline{\pi}_0(a) &= \pi_0^t(\gamma(a))\\
&= \pi_0^t(S(q^{h_\rho}aq^{-h_\rho}))\\
&= \pi_0^*(q^{h_\rho}aq^{-h_\rho})\\
\Rightarrow\ \ \overline{\pi}_0(a)_{ij} 
&= \pi_0^*( q^{h_\rho}aq^{-h_\rho})_{ij}\\
&= q^{(\rho,\varepsilon_j-\varepsilon_i)}\pi_0^*(a)_{ij}
\end{align*}
or
\begin{equation}
\pi_0^*(a)_{ij} = q^{(\rho,\varepsilon_i-\varepsilon_j)}\overline{\pi}_0(a)_{ij}.
\label{p26star}
\end{equation}

Now consider the dual vector $L$-operators:
\begin{align*}
(\pi_0^*\otimes \id)R &= (\pi_0^t\otimes \id)(S\otimes\id)R\\
&= (\pi_0^t\otimes \id)(\id\otimes S^{-1})R\\
&\stackrel{(\ref{p16stara})}{=} \sum_{i\leq j} e_{ij}\otimes S^{-1}(\hat{E}_{ij}), \\
(\pi_0^*\otimes \id)R^T &= (\pi_0^t\otimes \id)(S\otimes\id)R^T\\
&= (\pi_0^t\otimes \id)(\id\otimes S^{-1})R^T\\
&\stackrel{(\ref{p16starb})}{=} \sum_{i\leq j} e_{ji}\otimes S^{-1}(\hat{E}_{ji}). \\
\end{align*}
It makes more sense, however, to consider the physically more meaningful $L$-operators arising from
the (unitary) undeformed ``dual'' representation $\overline{\pi}_0$, referred to herein as the {\em
pseudo vector representation}. In view of equation (\ref{p26star}), this gives the following
$L$-operators:
\begin{align}
(\overline{\pi}_0\otimes\id)R &= \sum_{i\leq j}q^{(\rho,\varepsilon_j-\varepsilon_i)}e_{ij}\otimes
	S^{-1}(\hat{E}_{ij}), \label{p27stara}\\
(\overline{\pi}_0\otimes \id)R^T &= \sum_{i\leq j}
q^{(\rho,\varepsilon_i-\varepsilon_j)}e_{ji}\otimes S^{-1}(\hat{E}_{ji}).
	\label{p27starb}
\end{align}
The corresponding matrix 
$$
A = (q-q^{-1})^{-1}\left\{ I\otimes I - (\overline{\pi}_0\otimes \id)R^TR \right\}
$$
has entries
\begin{equation}
A_{ij} = (q-q^{-1})^{-1}\left\{ \delta_{ij} - \sum_{k\leq i\wedge j}
q^{(\rho,\varepsilon_j-\varepsilon_i)}S^{-1}(\hat{E}_{ik})S^{-1}(\hat{E}_{kj}) \right\}.
\label{p27star}
\end{equation}
It follows that the entries of the matrix $A$ transform according to
$$
aA_{ij} = (\overline{\pi}_0^*\otimes \overline{\pi}_0)\Delta(a_{(1)})_{\gamma i,\delta
j}A_{\gamma\delta}a_{(2)}.
$$
\underline{Notes}: (1) As seen in Appendix C, this agrees with the matrix $A$
of \cite{GLB1992}, although the
conventions used are quite different. \\
(2) The dual pseudo-vector representation is given by
$$
\overline{\pi}_0^*(a) = \pi_0^*(\gamma(a))^t = \pi_0(S(\gamma(a))),
$$
where 
$$
S(\gamma(a)) = q^{-h_\rho}\gamma^2(a)q^{h_\rho} =  q^{-h_\rho}aq^{h_\rho}.
$$
This implies that
$$
\overline{\pi}_0^*(a) = \pi_0(q^{-h_\rho}aq^{h_\rho}),
$$
so $\overline{\pi}_0^*$ is equivalent to the vector representation. Moreover,
$$
\overline{\pi}_0^*(a)_{ij} = q^{(\rho,\varepsilon_j - \varepsilon_i)}\pi_0(a)_{ij}.
$$

We have the following relations:
\begin{align}
q^{E_{kk}}A_{ij} &=
q^{(\varepsilon_i,\varepsilon_k)}q^{-(\varepsilon_k,\varepsilon_j)}A_{ij}q^{E_{kk}}
=q^{(\varepsilon_k,\varepsilon_i-\varepsilon_j)}A_{ij} q^{E_{kk}},\label{p28star}
\end{align}
\begin{align}
	& e_k A_{ij} = q^{(\varepsilon_k-\varepsilon_{k+1},\varepsilon_i-\varepsilon_j)/2}A_{ij}e_k +
(\overline{\pi}_0^*\otimes\overline{\pi}_0)(q^{h_k/2}\otimes e_k + e_k\otimes q^{-h_k/2})_{\gamma
i,\delta j} A_{\gamma\delta} q^{-h_k/2} \nonumber \\
	\Rightarrow \ \ &  e_kA_{ij} - q^{(\varepsilon_k-\varepsilon_{k+1},\varepsilon_i-\varepsilon_j)/2}A_{ij}e_k \nonumber\\
&\qquad = 
	q^{(\varepsilon_k-\varepsilon_{k+1},\varepsilon_i)/2}
	(-e_{k+1\ k})_{\delta j} A_{i\delta} q^{-h_k/2}
+ q^{(\rho,\varepsilon_i-\varepsilon_\gamma)}(e_{k\ k+1})_{\gamma
	i}q^{(\varepsilon_k-\varepsilon_{k+1},\varepsilon_j)/2}A_{\gamma j}q^{h_k/2}\nonumber\\
	&\qquad =
	-\delta_{jk}q^{(\varepsilon_k-\varepsilon_{k+1},\varepsilon_i)/2}A_{i\ k+1} q^{-h_k/2} + \delta_{i\
	k+1}q^{(\varepsilon_k-\varepsilon_{k+1},\varepsilon_j)/2 - 1}A_{kj}q^{-h_k/2}\label{p29star}
\end{align}
and similarly
\begin{align}
	& f_k A_{ij} = q^{(\varepsilon_k-\varepsilon_{k+1},\varepsilon_i-\varepsilon_j)/2}A_{ij} f_k
+ (\overline{\pi}_0^*\otimes \overline{\pi}_0)(q^{h_k/2}\otimes f_k + f_k\otimes q^{-h_k/2})_{\gamma
i,\delta j}A_{\gamma \delta}q^{-h_k/2} \nonumber\\
	\Rightarrow \ \ & f_k A_{ij} - q^{(\varepsilon_k-\varepsilon_{k+1},\varepsilon_i-\varepsilon_j)/2}A_{ij} f_k\nonumber\\
	& \qquad = q^{(\varepsilon_k-\varepsilon_{k+1},\varepsilon_i)/2}(-e_{k\ k+1})_{\delta j}A_{i\delta} q^{-h_k/2}
+ q^{(\rho,\varepsilon_i - \varepsilon_\gamma)}(e_{k+1\ k})_{\gamma
	i}q^{(\varepsilon_k-\varepsilon_{k+1},\varepsilon_j)/2}A_{\gamma j}q^{-h_k/2}\nonumber\\
	& \qquad = -q^{(\varepsilon_k-\varepsilon_{k+1},\varepsilon_i)/2}\delta_{j\ k+1} A_{ik} q^{-h_k/2}
	+ q^{(\varepsilon_k-\varepsilon_{k+1},\varepsilon_j)/2 + 1}\delta_{ki}A_{k+1\ j} q^{-h_k/2}. \label{p29starstar}
\end{align}

Acting on an irreducible $U_q(n)$ module $V(\Lambda)$, the matrix $A$ may be regarded
as an intertwining operator on $V_0^*\otimes V(\Lambda)$, where it is understood that the
representation afforded by $V_0^*$ is actually $\overline{\pi}_0$ (rather than $\pi_0^*$). In this
case, $A$ has eigenvalues
$$
a_r = \frac{1-q^{-2\alpha_r}}{q-q^{-1}},\ \ \alpha_r = \Lambda_r+n-r,
$$
with $\alpha_r$ the classical characteristic roots \cite{Gould1978}. Thus $A$ satisfies the polynomial identity
$$
\prod_{r=1}^n(A-a_r)=0.
$$
The corresponding projection operators
$$
P_r = \prod_{\ell\neq r}\left( \frac{A-a_\ell}{a_r-a_\ell} \right)
$$
project $V_0^*\otimes V(\Lambda)$ onto the submodule $V(\Lambda-\varepsilon_r).$

\noindent
\underline{Note}: 
Thus if $\{e_i^*\}$ is the basis for $V_0^*$ dual to the standard basis of $V_0$, then we are
defining
$$
ae_i^* = \overline{\pi}_0(a)_{ji}e_j^*,
$$
or 
$$
\langle a_i^*,e_j \rangle = \langle e_i^*,\gamma(a)e_j \rangle.
$$
Note, however, that if we set
$$
e_i^0 = q^{h_\rho}e_i^*
$$
then
\begin{align*}
ae_i^0 &= q^{h_\rho}(q^{-h_\rho}aq^{h_\rho})e_i^*\\
&=\overline{\pi}_0(q^{-h_\rho}aq^{h_\rho})_{ji} e_j^0 = \pi_0^*(a)_{ji}e_j^0.
\end{align*}
Therefore, the dual vector representation is obtained from $\overline{\pi}_0$ by a simple basis
transformation.
 

\subsection{Partitioning of matrix $A$}

The first $n-1$ rows and columns of the matrix $A$ gives the matrix $A_0$ of
$U_q(n-1)$. Setting $i=j=n$ into equations (\ref{p28star}), (\ref{p29star}) and
(\ref{p29starstar}) we obtain
$$
[e_k,A_{nn}] = [f_k,A_{nn}] = [q^{E_{kk}},A_{nn}] = 0
$$
so that $A_{nn}$ is an invariant of $U_q(n-1)$.

Setting $j=n$ into these equations, the entries of the last column, i.e.
$$
\overline{\psi}_i = A^i_n,\ \ 1\leq i<n
$$
satisfy 
$$
q^{E_{kk}}A_{in} = q^{(\varepsilon_k,\varepsilon_i)}A_{in}q^{E_{kk}},
$$
i.e.
$$
q^{E_{kk}}\overline{\psi}_i = q^{(\varepsilon_k,\varepsilon_i)}\overline{\psi}_iq^{E_{kk}},
$$
\begin{align*}
	e_k\overline{\psi}_i - q^{(\varepsilon_k-\varepsilon_{k+1},\varepsilon_i)/2}\overline{\psi}_ie_k &= \delta_{i\
k+1}q^{-1}\overline{\psi}_k q^{-h_k/2},\\
	f_k\overline{\psi}_i - q^{(\varepsilon_k-\varepsilon_{k+1},\varepsilon_i)/2}\overline{\psi}_i f_k &= \delta_{ik} q
\overline{\psi}_{k+1}q^{-h_k/2},
\end{align*}
and thus transform as a vector operator of rank $\overline{\pi}_0^*$ (i.e. a dual pseudo vector operator).

Setting $i=n$ into the same equations, the entries of the last row, i.e.
$$
\overline{\phi}_i = A_{ni},\ \ 1\leq i<n
$$
satisfy
$$
q^{E_{kk}}\overline{\phi}_i = q^{-(\varepsilon_k,\varepsilon_i)}\overline{\phi}_iq^{E_{kk}},\\
$$
\begin{align*}
	e_k\overline{\phi}_i - q^{-(\varepsilon_k-\varepsilon_{k+1},\varepsilon_i)/2}\overline{\phi}_ie_k &=
-\delta_{ik}\overline{\phi}_{k+1} q^{-h_k/2},\\
	f_k\overline{\phi}_i -
	q^{-(\varepsilon_k-\varepsilon_{k+1},\varepsilon_i)/2}\overline{\phi}_if_k
	& = -\delta_{i\ k+1}\overline{\phi}_k q^{-h_k/2},
\end{align*}
and thus transform as a pseudo vector operator (i.e. as the undeformed dual vector representation
$\overline{\pi}_0$).

Thus we have the following partitioning
$$
A = \left( \begin{array}{cc|c} A_0 &  & A_{in}\\
&&\\
\hline
 A_{ni} && A_{nn} \end{array} \right),
$$
with $A_0$ the characteristic matrix of $U_q(n-1)$, $A_{in}$ a $U_q(n-1)$
vector operator of rank $\overline{\pi}_0^*$, $A_{ni}$ a $U_q(n-1)$ pseudo vector
operator, and $A_{nn}$ a $U_q(n-1)$ invariant.

\noindent
\underline{Note}: The above transformation laws also apply to the matrix powers of $A$,
$\tilde{A}$. In particular,
$$
\left(A^m\right)_{nn},\ \ \left( \tilde{A}^m\right)_{nn}
$$
determine $U_q(n-1)$ invariants. One of our aims is to systematically determine the
eigenvalues of these invariants.


\section{On Invariants and Wigner coefficients} \label{Section5}

In view of Appendix B, Lemma \ref{lemmaBprime}, since the matrices $A$, $\tilde{A}$ transform as
tensor operators of type $\overline{\pi}_0^*\otimes \overline{\pi}_0$, $\pi_0^*\otimes \pi_0$
respectively, the following are invariants:
\begin{align*}
C_m &\equiv \mbox{tr}(\overline{\pi}_0\otimes\id)(q^{2h_\rho}\otimes I)\frac{\left(I -
R^TR\right)^m}{(q-q^{-1})^m}\\
&= \mbox{tr}\left( \overline{\pi}_0(q^{2h_\rho})A^m \right) \\
&= \mbox{tr}_q\left( A^m \right) = \sum_{i=1}^n q^{-(2\rho,\varepsilon_i)}A^m_{ii},
\end{align*}
\begin{align*}
\tilde{C}_m &\equiv \mbox{tr}\left( \pi_0(q^{2h_\rho}) \tilde{A}^m \right)\\
&= \sum_{i=1}^n q^{(2\rho,\varepsilon_i)}\tilde{A}^m_{ii},
\end{align*}
so $C_m$, $\tilde{C}_m$ belong to the centre of $ U_q(n)$.

In particular, we have the first order invariants
\begin{align*}
C_1 &= \sum_{i=1}^nq^{-(2\rho,\varepsilon_i)} A_{ii} = \sum_{i=1}^n q^{-(n+1-2i)} A_{ii},\\
\tilde{C}_1 &= \sum_{i=1}^n q^{(2\rho,\varepsilon_i})\tilde{A}_{ii} = \sum_{i=1}^n
q^{n+1-2i}\tilde{A}_{ii}. 
\end{align*}
On a maximal weight vector $v_+\in V(\Lambda)$ we observe that
$$
A_{ii} = (q-q^{-1})^{-1}\left\{ I - \sum_{k\leq i} S^{-1}(\hat{E}_{ik})S^{-1}(\hat{E}_{ki}) \right\}
$$
satisfies
$$
A_{ii}v_+ = \frac{1-q^{-2E_{ii}} }{q-q^{-1}} v_+ = \frac{1-q^{-2(\Lambda,\varepsilon_i)}}{q-q^{-1}}
v_+.
$$
Therefore the eigenvalue of $C_1$ on $V(\Lambda)$ is given by
\begin{align*}
\chi_\Lambda(C_1) &= \sum_{i=1}^n q^{-(2\rho,\varepsilon_i)}
\frac{1-q^{-2(\Lambda,\varepsilon_i)}}{q-q^{-1}}\\
&= \sum_{i=1}^n q^{-(\Lambda+2\rho,\varepsilon_i)} [(\Lambda,\varepsilon_i)]_q,
\end{align*}
where we have used the $q$-number notation
$$
[x]_q = \frac{q^x-q^{-x}}{q-q^{-1}}.
$$
Similarly
$$
\tilde{A}_{ii} = (q-q^{-1})^{-1}\left\{ I - \sum_{k\leq i} S^{-1}(\hat{E}_{ki})S(\hat{E}_{ik}) \right\}
$$
and so on a minimal weight vector $v_-\in V(\Lambda)$ of weight $\Lambda_-$, we have 
\begin{align*}
\tilde{A}_{ii}v_- &= (q-q^{-1})^{-1}\left\{ I - q^{-2E_{ii}} \right\} v_-\\
&= (q-q^{-1})\left( 1 - q^{-2(\Lambda_-,\varepsilon_i)} \right)v_-.
\end{align*}
Therefore, on $V(\Lambda)$ the invariant $\tilde{C}_1$ has eigenvalue
$$
\chi_\Lambda(\tilde{C}_1) = \sum_{i=1}^n
q^{(2\rho,\varepsilon_i)}\frac{1-q^{-2(\Lambda_-,\varepsilon_i)}}{q-q^{-1}}.
$$

Now if $\tau$ is the unique Weyl group element sending positive roots to negative roots (and thus
maximal weights to minimal weights), we have
\begin{align}
\chi_\Lambda(\tilde{C}_1) &= \sum_{i=1}^n
q^{(2\rho,\tau(\varepsilon_i))}\frac{1-q^{-2(\Lambda_-,\tau(\varepsilon_i))}}{q-q^{-1}}\nonumber\\
&= \sum_{i=1}^n q^{(2\tau(\rho),\varepsilon_i)}
\frac{1-q^{-2(\tau(\Lambda_-),\varepsilon_i)}}{q-q^{-1}}\nonumber\\
&= \sum_{i=1}^n q^{-(2\rho,\varepsilon_i)}
	\frac{1-q^{-2(\Lambda,\varepsilon_i)}}{q-q^{-1}} \nonumber\\
&= \sum_{i=1}^n q^{-(\Lambda+2\rho,\varepsilon_i)} [(\Lambda,\varepsilon_i)]_q.
	\label{p35star}
\end{align}
Thus $C_1$ and $\tilde{C}_1$ have the same eigenvalues.

Finally we have the spectral decompositions
$$
A^m = \sum_{r=1}^n a_r^m P_r,\ \ \tilde{A}^m = \sum_{r=1}^n \tilde{a}_r^m \tilde{P}_r,
$$
\begin{align*}
\Rightarrow \ \ C_m &= \mbox{tr}\left( \overline{\pi}_0(q^{2h_\rho})A^m \right) \equiv \tau_q(A^m) =
\sum_{r=1}^n a_r^m \tau_q(P_r),\\
\tilde{C}_m &= \mbox{tr}\left( \pi_0(q^{2h_\rho})\tilde{A}^m \right) \equiv \tau_q(\tilde{A}^m) =
\sum_{r=1}^n \tilde{a}_r^m \tau_q(\tilde{P}_r).
\end{align*}
We have
\begin{align*}
D_q[\Lambda]\tau_q(P_r) &= \mbox{tr}_{V_0^*\otimes V(\Lambda)}[(q^{2h_\rho}\otimes q^{2h_\rho})P_r] =
D_q[\Lambda-\varepsilon_r],\\
D_q[\Lambda]\tau_q(\tilde{P}_r) &= \mbox{tr}_{V_0\otimes V(\Lambda)}[(q^{2h_\rho}\otimes
q^{2h_\rho})\tilde{P}_r] = D_q[\Lambda+\varepsilon_r],
\end{align*}
where
$$
D_q[\Lambda] = \prod_{\alpha>0}\frac{q^{(\Lambda+\rho,\alpha)} -
q^{-(\Lambda+\rho,\alpha)}}{q^{(\rho,\alpha)} - q^{-(\rho,\alpha)}} =
\prod_{\alpha>0}\frac{[(\Lambda+\rho,\alpha)]_q}{[(\rho,\alpha)]_q} 
$$
is the usual $q$-dimension.
Thus we obtain
\begin{align*}
\chi_\Lambda(C_m) &= \sum_{r=1}^na_r^m\prod_{\alpha>0}\frac{[(\Lambda+\rho -
\varepsilon_r,\alpha)]_q}{[(\Lambda+\rho,\alpha)]_q},\\
\chi_\Lambda(\tilde{C}_m) &= \sum_{r=1}^n\tilde{a}_r^m\prod_{\alpha>0}\frac{[(\Lambda+\rho +
\varepsilon_r,\alpha)]_q}{[(\Lambda+\rho,\alpha)]_q}.
\end{align*}
Our aim, following \cite{GLB1992,Gould1992}, is to utilise the characteristic identities for $U_q(n)$
and $U_q(n-1)$ to evaluate the $U_q(n-1)$ invariants
$$
\left( A^m\right)_{nn} = \sum_{k=1}^n a_k^n\omega_k,\ \ \left( \tilde{A}^m \right)_{nn}
= \sum_{k=1}^n\tilde{a}_k^m\tilde{\omega}_k,
$$
where
$$
\omega_k = \left( P_k \right)_{nn},\ \ \tilde{\omega}_k = \left( \tilde{P}_k \right)_{nn}.
$$
These latter invariants are of particular interest since their eigenvalues determine the squared
RWCs
$$
\left\| \left\langle \begin{array}{c} \Lambda-\varepsilon_k\\ \Lambda_0 \end{array} \right| 
\left| \begin{array}{c} \overline{\varepsilon_1}\\ \dot{0}\end{array} ; \begin{array}{c}
\Lambda\\ \Lambda_0\end{array} \right\rangle \right\|^2,\ \ 
\left\| \left\langle \begin{array}{c} \Lambda+\varepsilon_k\\ \Lambda_0 \end{array} \right| 
\left| \begin{array}{c} \overline{\varepsilon_1}\\ \dot{0}\end{array} ; \begin{array}{c}
\Lambda\\ \Lambda_0\end{array} \right\rangle \right\|^2
$$ 
respectively, where $\Lambda_0$ is the highest weight of an irreducible $U_q(n-1)$
submodule of $V(\Lambda)$, and $\dot{0}$ denotes the $U_q(n-1)$ weight
$(0,0,\ldots,0)$.


\subsection{Evaluation of $\omega_k,$ $\tilde{\omega}_k$}

Using the $U_q(n)$ characteristic identity we have
$$
(P_kA)^n_i = a_k (P_k)^n_i
$$
which may be expanded to give
$$
a_k (P_k)_{ni} = \omega_k\overline{\phi}_i + (P_k)_{nj}(A_0)_{ji},
$$
with $A_0$ the $U_q(n-1)$ characteristic matrix and $\overline{\phi}_i=A_{ni}$
(pseudo vector operator). Therefore we have
$$
\omega_k\overline{\phi}_i = (P_k)_{nj}(A_0-a_k)_{ji}
$$
$$
\Rightarrow \ \ \sum_r \omega_k\overline{\phi}[r]_i(a_k-a_{0r})^{-1} = (P_k)_{ni}
$$
where $a_{0r}$ are the $U_q(n-1)$ characteristic roots.

Now using the shift property (see Appendix B)
$$
\overline{\phi}[r]_ia_{0r} = (q^{-2}a_{0r}+q^{-1})\overline{\phi}[r]_i,
$$
the above becomes 
\begin{equation}
\sum_r \omega_k(a_k - q^{-2}a_{0r}-q^{-1})^{-1}\overline{\phi}[r]_i = (P_k)_{ni}.
\label{p37star}
\end{equation}
Using 
$$
\sum_{k=1}^n(P_k)_{ni}=\delta_{ni}
$$
we arrive at the equations
\begin{align*}
& \sum_{k=1}^n \omega_k(a_k - q^{-2}a_{0r} - q^{-1})^{-1} = 0,\ \ 1\leq r<n,\\
& \sum_{k=1}^n \omega_k = 1.
\end{align*}
Following \cite{Gould1992}, these equations uniquely determine the $\omega_k$ and yield the solution
\begin{equation}
\omega_k = \prod_{r=1}^{n-1}(a_k - q^{-2}a_{0r} - q^{-1})\prod_{\ell\neq k}^n(a_k-a_\ell)^{-1}.
\label{p38star}
\end{equation}
Similarly applying the identity for the matrix $\tilde{A}$ we have
$$
\tilde{a}_k(\tilde{P}_k)_{in} = (\tilde{A}\tilde{P}_k)_{in} = (\tilde{A})_{ij} (\tilde{P}_k)_{jn} +
\phi_i\tilde{\omega}_k,
$$
where $\phi_i = \tilde{A}_{in}$ (dual vector operator). Therefore
\begin{equation}
\phi_i\overline{\omega}_k = (\tilde{a}_k - A_0)_{ij}(\tilde{P}_k)_{jn},
\label{p38starstar}
\end{equation}
$$
\Rightarrow \ \ \sum_{r=1}(\tilde{a}_k-\tilde{a}_{0r})^{-1}\phi[r]_i\tilde{\omega}_k = (\tilde{P}_k)_{jn}.
$$
Here we note that, from Appendix B,
\begin{align*}
&\tilde{a}_{0r}\phi[r]_i = \phi[r]_i(q^2\tilde{a}_{0r} - q)\\
\Rightarrow \ \ & \sum_{r=1}^{n-1}\phi[r]_i(\tilde{a}_k - q^2\tilde{a}_{0r} +
q)^{-1}\tilde{\omega}_k = (\tilde{P}_k)_{in}. 
\end{align*}
From this we obtain the equations
\begin{align*}
	& \sum_{k=1}^n (\tilde{a}_k - q^{2}\tilde{a}_{0r} + q)^{-1}\tilde{\omega}_k = 0,\ \ 1\leq r<n,\\
& \sum_{k=1}^n \tilde{\omega}_k = 1,
\end{align*}
which yield the unique solution
\begin{equation}
\tilde{\omega}_k = \prod_{r=1}^{n-1}(\tilde{a}_k - q^2\tilde{a}_{0r} + q)\prod_{\ell\neq
k}^n(\tilde{a}_k-\tilde{a}_\ell)^{-1}.
\label{p39star}
\end{equation}

Alternatively,
$$
(\tilde{P}_k\tilde{A})_{ni} = \tilde{a}_k(\tilde{P}_k)_{ni} = (\tilde{P}_k)_{nj}(\tilde{A}_0)_{ji} +
\tilde{\omega}_k\psi_i,
$$
with $\psi_i=\tilde{A}_{ni}$ a vector operator. Therefore
$$
\tilde{\omega}_k\psi_i = (\tilde{P}_k)_{nj}(\tilde{a}_k - \tilde{A}_0)_{ji}
$$
\begin{align}
(\tilde{P}_k)_{nj} &= \sum_{r=1}^{n-1}\tilde{\omega}_k\psi[r]_i(\tilde{a}_k - \tilde{a}_{0r})^{-1}
\nonumber \\
&= \sum_{r=1}^{n-1}\tilde{\omega}_k(\tilde{a}_k - q^2\tilde{a}_{0r} + q)^{-1}\psi[r]_i
\label{p39starstar}
\end{align}
where we have used (c.f. Appendix B)
$$
\psi[r]_i\tilde{a}_{0r} = (q^2\tilde{a}_{0r} - q)\psi[r]_i.
$$
this yields exactly the same equation to solve for the $\tilde{\omega}_k$ as above.


\subsection{Reduced matrix elements}

Below we adopt the following notation:
\begin{align*}
\tilde{\phi}_i &= \tilde{A}_{in} \mbox{ -- dual vector operator of $U_q(n-1)$ (matrix
$\tilde{A}$ acts on {\em left})},\\
\tilde{\psi}_i &= \tilde{A}_{ni} \mbox{ -- vector operator of $U_q(n-1)$ (matrix
$\tilde{A}$ acts on {\em right})},\\ 
\overline{\phi}_i &= {A}_{ni} \mbox{ -- pseudo vector operator of $U_q(n-1)$ (matrix
$\tilde{A}$ acts on {\em right})},\\ 
\overline{\psi}_i &= A_{in} \mbox{ -- dual pseudo vector operator of $U_q(n-1)$ (matrix
$\tilde{A}$ acts on {\em left})}.
\end{align*}
\underline{Note}: With this notation, letting $\rho_0$ denote the half sum of positive roots
from $U_q(n-1)$, observe that 
$$
\psi_i = q^{-(\rho_0,\varepsilon_i)} \overline{\psi}_i
$$
is a vector operator and
$$
\overline{\phi}_i = q^{-(\rho_0,\varepsilon_i)}\overline{\phi}_i
$$
is a dual vector operator.

Now from equation (\ref{p39starstar}) we have
$$
(\tilde{P}_k)_{nj} = \sum_{r=1}^{n-1}\tilde{\omega}_k(\tilde{a}_k - q^2\tilde{a}_{0r} +
q)^{-1}\tilde{\psi}[r]_i.
$$
Therefore we look for solutions $\tilde{\gamma}_{rk}$ to equations
\begin{align}
& \sum_{k=1}^n \tilde{\gamma}_{rk}\tilde{\omega}_k (\tilde{a}_k - q^2\tilde{a}_{0\ell}+q)^{-1} =
\delta_{r\ell}, \label{p41stara}\\
& \sum_{k=1}^n \tilde{\gamma}_{rk}\tilde{\omega}_k = 0. \label{p41starb}
\end{align}
This yields the unique solution
$$
\tilde{\gamma}_{rk} = \tilde{\gamma}_r(\tilde{a}_k - q^2\tilde{a}_{0r} + q)^{-1},
$$
where
\begin{align}
\tilde{\gamma}_r &= (-1)^{n-1}\frac{\prod_{k=1}^n(\tilde{a}_k -
q^2\tilde{a}_{0r}+q)}{\prod_{\ell\neq r}^{n-1}(q^2\tilde{a}_{0r}-q^2\tilde{a}_{0\ell})}
\label{p41starstar}\\
&= (-1)^{n-1} q^{4-n}\frac{\prod_{k=1}^n(q^{-1}\tilde{a}_k -
q\tilde{a}_{0r}+1)}{\prod_{\ell\neq r}^{n-1}(\tilde{a}_{0r}-\tilde{a}_{0\ell})}.
\nonumber
\end{align}
These invariants have an interesting interpretation. From equations
(\ref{p41stara}),(\ref{p41starb}) above, we obtain
\begin{align*}
\sum_{k=1}^n\tilde{\gamma}_{rk}(\tilde{P}_k)_{ni} &=
\sum_{\ell=1}^{n-1}\sum_{k=1}^n\tilde{\gamma}_{rk}\tilde{\omega}_k(\tilde{a}_k-q^2\tilde{a}_{0\ell}+q)^{-1}\tilde{\psi}[\ell]_i\\
&= \tilde{\psi}[r]_i. 
\end{align*}
Therefore (summation on $i$)
\begin{align*}
\tilde{\psi}[r]_i\tilde{\phi}[r]_i &= \tilde{\psi}[r]_i\tilde{\phi}_i\\
&= \tilde{\psi}[r]_i\tilde{A}_{in}\\
&= \sum_{k=1}^n\tilde{\gamma}_{rk}(\tilde{P}_k)_{ni}\tilde{A}_{in}\\
&= \sum_{k=1}^n \tilde{\gamma}_{rk} \left( \tilde{a}_k\tilde{\omega}_k -
\tilde{\omega}_k\tilde{A}_{nn} \right)\\
&= \tilde{\gamma}_r\sum_{k=1}^n(\tilde{a}_k - q^2\tilde{a}_{0r}+q)^{-1}\tilde{a}_k\tilde{\omega}_k\\
&= \tilde{\gamma}_r \sum_{k=1}^n\left( \tilde{\omega}_k + (q^2\tilde{a}_{0r} -
q)(\tilde{a}_k-q^2\tilde{a}_{0r}+q)^{-1}\tilde{\omega}_k \right)\\
&= \tilde{\gamma}_r.
\end{align*}
Thus we can think of $\tilde{\gamma}_r$ as determining the ``length'' of the
shift vector
$\tilde{\phi}[r]_i.$ 

This gives the projection property
\begin{equation}
\tilde{\phi}[r]_i (\tilde{\gamma}_r)^{-1}\tilde{\psi}[r]_j = (\tilde{P}_r)_{ij},
\label{p42star}
\end{equation}
which generalises the projection property for normal vectors. We note that
\begin{align*}
\tilde{\gamma}_r\tilde{\psi}[r]_i &= (-1)^{n-1}\frac{\prod_{k=1}^n(\tilde{a}_k -
q^2\tilde{a}_{0r}+q)}{\prod_{\ell\neq r}(q^2\tilde{a}_{0r}-q^2\tilde{a}_{0\ell})} \tilde{\psi}[r]_i\\
&= \tilde{\psi}[r]_i\tilde{\mu}_r,
\end{align*}
where
\begin{equation}
\tilde{\mu}_r = (-1)^{n-1}\frac{\prod_{k=1}^n(\tilde{a}_k - \tilde{a}_{0r})}{\prod_{\ell\neq
r}(\tilde{a}_{0r}-q^2\tilde{a}_{0\ell}+q)}
\label{p43star}
\end{equation}
and where we have used (c.f. Appendix B)
\begin{align*}
(q^2\tilde{a}_{0r}-q)\tilde{\psi}[r]_i &= \tilde{\psi}[r]_i\tilde{a}_{0r},\\
	\tilde{a}_{0r}\tilde{\psi}[r]_i &= \tilde{\psi}[r]_i(q^{-2}\tilde{a}_{0r} + q^{-1}).
\end{align*}
Thus equation (\ref{p42star}) may be rearranged to give
\begin{equation}
\tilde{\phi}[r]_i\tilde{\psi}[r]_j = \tilde{\mu}_r(\tilde{P}_r)_{ij}.
\label{p43starstar}
\end{equation}
The invariants $\tilde{\mu}_r$ are of interest since they determine the squared reduced matrix
elements (RMEs) of the vector operator $\tilde{\psi}[r]_i$.

\noindent
\underline{Note}: It is worth observing that
$$
\tilde{\gamma}_r(\Lambda,\Lambda_0+\varepsilon_r) = \tilde{\mu}_r(\Lambda,\Lambda_0)
$$
which shows the relation between the two invariants.

Similarly using the identity for the matrix $A$, we have from equation (\ref{p37star})
$$
(P_k)_{ni} = \sum_{\ell=1}^{n-1}\omega_k(a_k - q^{-2}a_{0\ell}-q^{-1})^{-1}\overline{\phi}[\ell]_i.
$$
Now we consider the solution $\gamma_{rk}$ to
\begin{align*}
& \sum_{k=1}^n\gamma_{rk}\omega_k(a_k-q^{-2}a_{0\ell}-q^{-1})^{-1} = \delta_{r\ell},\\
& \sum_{k=1}^n\gamma_{rk}\omega_k=0,
\end{align*}
which yields the solution
$$
\gamma_{rk} = (a_k - q^{-2}a_{0r}-q^{-1})^{-1}\gamma_r,
$$
where
\begin{align*}
\gamma_r &= (-1)^{n-1}\frac{\prod_{k=1}^n(a_k-q^{-2}a_{0r}-q^{-1})}{\prod_{\ell\neq r}(q^{-2}a_{0r} -
q^{-2}a_{0\ell})}\\
&= (-1)^{n-1}q^{n-4}\frac{\prod_{k=1}^n(qa_k-q^{-1}a_{0r} - 1)}{\prod_{\ell\neq r}(a_{0r}-a_{0\ell})}.
\end{align*}
In this case we may write 
$$
\overline{\phi}[r]_i = \sum_k\gamma_{rk}(P_k)_{ni}
$$
so that, multiplying on the right by $\overline{\psi}_i=A_{in}$, we obtain (summation on $i$)
$$
\overline{\phi}[r]_i\overline{\psi}[r]_i = \sum_k\gamma_{rk}(P_k)_{ni}A_{in} = \gamma_r.
$$
We note that 
$$
\gamma_r\overline{\phi}[r]_i = \overline{\phi}[r]_i\mu_r
$$
where now
\begin{align}
\mu_r(\Lambda,\Lambda_0) &= \gamma_r(\Lambda,\Lambda_0 - \varepsilon_r)\nonumber\\
&= (-1)^{n-1}\frac{\prod_{k=1}^n(a_k-a_{0r})}{\prod_{\ell\neq r}(a_{0r}-q^{-2}a_{0\ell}-q^{-1})}.
\label{45star}
\end{align}
Thus in this case we have the spectral decomposition
\begin{equation}
\overline{\psi}[r]_i(\gamma_r)^{-1}\overline{\phi}[r]_j = (P_r)_{ij} \label{p45starstara}
\end{equation}
or
\begin{equation}
\overline{\psi}[r]_i\overline{\phi}[r]_j = \mu_r(P_r)_{ij}. \label{p45starstarb}
\end{equation}


\subsection{Squared reduced Wigner coefficients}

Using the identity satisfied by the matrix $\tilde{A}$ we have (c.f. equation
(\ref{p38starstar}))
$$
\tilde{A}_{in}(\tilde{P}_k)_{nj} = (\tilde{a}_k - \tilde{A}_0)_{i\ell}(\tilde{P}_k)_{\ell j},
$$
or
$$
\tilde{\phi}_i(\tilde{P}_k)_{nj} = (\tilde{a}_k - \tilde{A}_0)_{i\ell}(\tilde{P}_k)_{\ell j}.
$$
Therefore, multiplying on the left by the subalgebra projector $\tilde{P}_{0r}$
($U_q(n-1)$ analogue of $\tilde{P}_k$) gives
\begin{align*}
\tilde{\phi}[r]_i(\tilde{P}_k)_{nj} &= (\tilde{a}_k -
\tilde{a}_{0r})(\tilde{P}_{0r})_{i\ell}(\tilde{P}_k)_{\ell j} \\
\Rightarrow \ \ (\tilde{P}_{0r}\tilde{P}_k)_{ij} &= (\tilde{a}_k -
\tilde{a}_{0r})^{-1}\tilde{\phi}[r]_i(\tilde{P}_k)_{nj}.
\end{align*}
Therefore from equation (\ref{p39starstar}) we obtain
$$
(\tilde{P}_{0r}\tilde{P}_k\tilde{P}_{0r})_{ij} = (\tilde{a}_k -
\tilde{a}_{0r})^{-1}\tilde{\phi}[r]_i \tilde{\omega}_k (\tilde{a}_k - q^2\tilde{a}_{0r} + q)^{-1}
\tilde{\psi}[r]_j.
$$
Now using the fact that the expression
$$
\tilde{\omega}_k(\tilde{a}_k - q^2\tilde{a}_{0r}+q)^{-1}
$$
is independent of $\tilde{a}_{0r}$ (c.f. formula (\ref{p39star}) we obtain
\begin{align*}
(\tilde{P}_{0r}\tilde{P}_k\tilde{P}_{0r})_{ij} &=
\tilde{\omega}_k(\tilde{a}_k-\tilde{a}_{0r})^{-1}(\tilde{a}_k -
q^2\tilde{a}_{0r}+q)^{-1}\tilde{\phi}[r]_i\tilde{\psi}[r]_j \\
& \stackrel{(\ref{p43starstar})}{=}
(\tilde{a}_k-\tilde{a}_{0r})^{-1}(\tilde{a}_k -
q^2\tilde{a}_{0r}+q)^{-1}\tilde{\omega}_k \tilde{\mu}_r(\tilde{P}_{0r})_{ij}.
\end{align*}
Thus we arrive at
\begin{equation}
(\tilde{P}_{0r}\tilde{P}_k\tilde{P}_{0r})_{ij} = \tilde{\omega}_{kr}(\tilde{P}_{0r})_{ij}
\label{p46star}
\end{equation}
where
$$
\tilde{\omega}_{kr}= (\tilde{a}_k-\tilde{a}_{0r})^{-1}(\tilde{a}_k -
q^2\tilde{a}_{0r}+q)^{-1}\tilde{\omega}_k \tilde{\mu}_r.
$$
These determine $U_q(n-1)$ invariant operators whose eigenvalues are the squared RWCs 
$$
\left| 
\left.
\left\langle 
  \begin{array}{c}
    \Lambda+\varepsilon_k\\
    \Lambda_0+\varepsilon_{0r}
  \end{array}
\right.
\right|
\left|
 \begin{array}{c}
    \varepsilon_1 \\
    \varepsilon_{01}
  \end{array}
\right.
;
\left.
  \begin{array}{c}
    \Lambda\\
    \Lambda_0
  \end{array}
\right\rangle
\right|^2
$$
where $\Lambda_0$ is the highest weight of a $U_q(n-1)$ submodule of $V(\Lambda)$.

Similarly we have
\begin{align*}
A_{in}(P_k)_{nj} + (A_0)_{i\ell}(P_k)_{\ell j} &= a_k(P_k)_{ij}\\
\Rightarrow \ \ \overline{\psi}_i(P_k)_{nj} &= (a_k-A_0)_{i\ell}(P_k)_{\ell j}.
\end{align*}
Now multiplying on the left by $U_q(n-1)$ projector $P_{0r}$ gives
\begin{align*}
\overline{\psi}[r]_i(P_k)_{nj} &= (a_k-a_{0r})(P_{0r}P_k)_{ij}\\
	\Rightarrow \ \ (P_{0r}P_k)_{ij} &= (a_k - a_{0r})^{-1}\overline{\psi}[r]_i(P_k)_{nj}.
\end{align*}
Therefore, from equation (\ref{p37star}), we obtain
\begin{align*}
({P}_{0r}{P}_k{P}_{0r})_{ij} &= (a_k -
a_{0r})^{-1}\overline{\psi}[r]_i\omega_k(a_k-q^{-2}a_{0r}-q^{-1})^{-1}\overline{\phi}[r]_j\\
&\stackrel{(\ref{p45starstarb})}{=} 
(a_k-a_{0r})^{-1}(a_k - q^{-2}a_{0r} - q^{-1})^{-1}\omega_k\mu_r(P_{0r})_{ij}.
\end{align*}
Therefore, in this case we obtain
\begin{equation}
({P}_{0r}{P}_k{P}_{0r})_{ij} =\omega_{kr}(P_{0r})_{ij},
\label{p47star}
\end{equation}
where
$$
\omega_{kr} = (a_k-a_{0r})^{-1}(a_k - q^{-2}a_{0r}-q^{-1})^{-1}\omega_k\mu_r.
$$
These invariants determine the squared pseudo vector RWCs
$$
\left| 
\left.
\left\langle 
  \begin{array}{c}
    \Lambda-\varepsilon_k\\
    \Lambda_0-\varepsilon_{0r}
  \end{array}
\right.
\right|
\left|
 \begin{array}{c}
    \overline{\varepsilon}_1 \\
    \overline{\varepsilon}_{01}
  \end{array}
\right.
;
\left.
  \begin{array}{c}
    \Lambda\\
    \Lambda_0
  \end{array}
\right\rangle
\right|^2
$$
where $\overline{\varepsilon_1} = -\varepsilon_n$ is the
highest weight of the pseudo vector module, and similarly for $\overline{\varepsilon}_{01} =
-\varepsilon_{0\ n-1}$.


\subsection{Evaluation of Wigner coefficients}

Recall \cite{Gould1992} that a (pseudo) vector Wigner coefficient (WC) is expressible as a product of RWCs for
each quantum group in the canonical subalgebra chain
$$
U_q(n)\supset U_q(n-1)\supset \cdots\supset
U_q(2)\supset U_q(1).
$$
It thus suffices to evaluate $U_q(n):U_q(n-1)$ (pseudo) vector RWCs
(also known as {\em isoscalar factors}).

For the vector RWCs, we have two types, namely
$$
\left.
\left\langle 
  \begin{array}{c}
    \Lambda+\varepsilon_k\\
    \Lambda_0
  \end{array}
\right.
\right|
\left|
 \begin{array}{c}
    {\varepsilon}_1 \\
    \dot{0}
  \end{array}
\right.
;
\left.
  \begin{array}{c}
    \Lambda\\
    \Lambda_0
  \end{array}
\right\rangle,\ \ 
\left.
\left\langle 
  \begin{array}{c}
    \Lambda+\varepsilon_k\\
    \Lambda_0 + \varepsilon_{0r}
  \end{array}
\right.
\right|
\left|
 \begin{array}{c}
    {\varepsilon}_1 \\
    \varepsilon_{01}
  \end{array}
\right.
;
\left.
  \begin{array}{c}
    \Lambda\\
    \Lambda_0
  \end{array}
\right\rangle
$$
corresponding to the two irreducible $U_q(n-1)$ submodules of the
$U_q(n)$ vector module $V=V(\varepsilon_1)$. We have seen that the invariants
$\tilde{\omega}_k$, $\tilde{\omega}_{kr}$ determine the absolute value squared of these RWCs
respectively, so it remains to determine the phases. The latter are determined by the phases $\pm 1$
of the corresponding classical RWCs obtained in the limit $q\rightarrow 1$. By this means, we obtain
\begin{align*}
\left.
\left\langle 
  \begin{array}{c}
    \Lambda+\varepsilon_k\\
    \Lambda_0
  \end{array}
\right.
\right|
\left|
 \begin{array}{c}
    {\varepsilon}_1 \\
    \dot{0}
  \end{array}
\right.
;
\left.
  \begin{array}{c}
    \Lambda\\
    \Lambda_0
  \end{array}
\right\rangle
 &= \tilde{\omega}_k^{1/2},
\\
\left.
\left\langle 
  \begin{array}{c}
    \Lambda+\varepsilon_k\\
    \Lambda_0 + \varepsilon_{0r}
  \end{array}
\right.
\right|
\left|
 \begin{array}{c}
    {\varepsilon}_1 \\
    \varepsilon_{01}
  \end{array}
\right.
;
\left.
  \begin{array}{c}
    \Lambda\\
    \Lambda_0
  \end{array}
\right\rangle
= S(r-k)\tilde{\omega}_{kr}^{1/2}
\end{align*}
where
$$
S(x) = \mbox{sign}(x), \mbox{ with } S(0)=1.
$$
Similarly we obtain
\begin{align*}
\left.
\left\langle 
  \begin{array}{c}
    \Lambda-\varepsilon_k\\
    \Lambda_0
  \end{array}
\right.
\right|
\left|
 \begin{array}{c}
    \overline{\varepsilon}_1 \\
    \dot{0}
  \end{array}
\right.
;
\left.
  \begin{array}{c}
    \Lambda\\
    \Lambda_0
  \end{array}
\right\rangle
 &= {\omega}_k^{1/2},
\\
\left.
\left\langle 
  \begin{array}{c}
    \Lambda-\varepsilon_k\\
    \Lambda_0 - \varepsilon_{0r}
  \end{array}
\right.
\right|
\left|
 \begin{array}{c}
    \overline{\varepsilon}_1 \\
    \overline{\varepsilon}_{01}
  \end{array}
\right.
;
\left.
  \begin{array}{c}
    \Lambda\\
    \Lambda_0
  \end{array}
\right\rangle
= S(r-k)\tilde{\omega}_{kr}^{1/2}
\end{align*}
\underline{Note}: Here we adopt the convention that the square root is that with positive real part.


\subsection{Alternative expressions}

Following Appendix D we may write
\begin{align*}
a_k-q^{-2}a_{0r}-q^{-1} &= q^{-(\alpha_k+\alpha_{0r}+1)}[\alpha_k-\alpha_{0r}-1]_q,\\
a_k - a_\ell &= q^{-(\alpha_k+\alpha_\ell)}[\alpha_k-\alpha_\ell]_q\\
\Rightarrow \ \ \omega_k &= \frac{\prod_{r=1}^{n-1}(a_k-q^{-2}a_{0r}-q^{-1})}{\prod_{\ell\neq
k}^n(a_k-a_\ell)}\\
&= \frac{\prod_{r=1}^{n-1}q^{-(\alpha_k+\alpha_{0r}+1)}[\alpha_k-\alpha_{0r}-1]_q}{\prod_{\ell\neq
k}^n q^{-(\alpha_k+\alpha_\ell)}[\alpha_k-\alpha_\ell]_q}\\
&= q^{\xi_k}\prod_{r=1}^{n-1}\frac{[\alpha_k-\alpha_{0r}-1]_q}{[\alpha_k-\alpha_\ell]_q},
\end{align*}
where 
\begin{align*}
\xi_k &= \sum_{\ell\neq k}^n\alpha_\ell - \sum_{r=1}^{n-1}(\alpha_{0r}+1)\\
&=\sum_{\ell=1}^n\Lambda_\ell - \sum_{r=1}^{n-1}\Lambda_{0r} - \alpha_k.
\end{align*}
\underline{Note}: the first two terms on the right hand side of the previous expression actually
determine the $n$th component of the weight of the semi-maximal state
$$
\left|
 \begin{array}{c}
    \Lambda \\
    \Lambda_0 \\
    \mbox{[max]}
 \end{array}
\right\rangle,
$$
which has weight $\Lambda_0 + \theta(\Lambda,\Lambda_0)\varepsilon_n$, where
$$
\theta(\Lambda,\Lambda_0) = \sum_{\ell=1}^n\Lambda_\ell - \sum_{r=1}^{n-1}\Lambda_{0r}.
$$

Similarly, for $\tilde{\omega}_k$ we have
\begin{align*}
\tilde{\omega}_k &= \prod_{r=1}^{n-1}(\tilde{a}_k - q^2\tilde{a}_{0r}+q)\prod_{\ell\neq
k}(\tilde{a}_k - \tilde{a}_\ell)^{-1}\\
&= \frac{ \prod_{r=1}^{n-1}q^{-(\overline{\alpha}_k + \overline{\alpha}_{0r}-1)}[\overline{\alpha}_k
- \overline{\alpha}_{0r}+1]_q }{ \prod_{\ell\neq k}^n
q^{-(\overline{\alpha}_k+\overline{\alpha}_\ell)} [\overline{\alpha}_k - \overline{\alpha}_\ell]_q
}\\
&= q^{\tilde{\xi}_k}\frac{ \prod_{r=1}^{n-1}[\overline{\alpha}_k
- \overline{\alpha}_{0r}+1]_q }{ \prod_{\ell\neq k}^n [\overline{\alpha}_k - \overline{\alpha}_\ell]_q }
\end{align*}
where 
\begin{align}
\tilde{\xi}_k &= \sum_{\ell\neq k}^n\overline{\alpha}_\ell -
\sum_{r=1}^{n-1}(\overline{\alpha}_{0r}-1)\nonumber\\
	&= \theta(\Lambda,\Lambda_0) - \overline{\alpha}_k\label{xitilde}\\
&= \theta(\Lambda,\Lambda_0) - \alpha_k + n-1 = \xi_k+n-1.\nonumber
\end{align}

Similarly, for the squared RMEs we have 
$$
\tilde{\mu}_r = (-1)^{n-1}\frac{ \prod_{k=1}^n(\tilde{a}_k - \tilde{a}_{0r}) }{ \prod_{\ell\neq
r}^{n-1}(\tilde{a}_{0r} - q^2\tilde{a}_{0\ell} + q) }.
$$
Now we use
\begin{align*}
\tilde{a}_k - \tilde{a}_{0r} &= q^{-(\overline{\alpha}_{0r} + \overline{\alpha}_k)
}[\overline{\alpha}_k - \overline{\alpha}_{0r}]_q,\\
\tilde{a}_{0r} - q^2\tilde{a}_{0\ell}+q &= q^{-(\overline{\alpha}_{0r} +
\overline{\alpha}_{0\ell}-1)}[\overline{\alpha}_{0r} - \overline{\alpha}_{0\ell}+1]_q\\
\Rightarrow \ \ \tilde{\mu}_r &= (-1)^{n-1}
q^{\tilde{\eta}_r}\frac{\prod_{k=1}^n[\overline{\alpha}_k -
\overline{\alpha}_{0r}]_q}{\prod_{\ell\neq r}^{n-1}[\overline{\alpha}_{0r} -
\overline{\alpha}_{0\ell}+1]_q },
\end{align*}
where
\begin{align}
\tilde{\eta}_r &= \sum_{\ell\neq r}^{n-1}(\overline{\alpha}_{0r} + \overline{\alpha}_{0\ell} - 1) -
\sum_{k=1}^n(\overline{\alpha}_{0r}+\overline{\alpha}_k)\nonumber\\
&= -2\overline{\alpha}_{0r} + \sum_{\ell\neq r}(\overline{\alpha}_{0\ell}-1) -
\sum_{k=1}^n\overline{\alpha}_k\nonumber\\
&= -3 \overline{\alpha}_{0r}+1+\sum_{\ell=1}^{n-1}\Lambda_{0\ell} -
	\sum_{k=1}^n\Lambda_k\label{etatilde}\\
&= -3\overline{\alpha}_{0r}+1-\theta(\Lambda,\Lambda_0).\nonumber
\end{align}
Also,
\begin{align*}
\mu_r &= (-1)^{n-1}\frac{ \prod_{k=1}^n(a_k - a_{0r}) }{ \prod_{\ell\neq r}^{n-1}(a_{0r} -
q^{-2}a_{0\ell} - q^{-1})}\\
&= (-1)^{n-1} \frac{ \prod_{k=1}^nq^{-(\alpha_{0r} + \alpha_k)}[\alpha_k - \alpha_{0r}]_q }{
\prod_{\ell\neq r}^{n-1} q^{-(\alpha_{0r}+\alpha_{0\ell}+1)}[\alpha_{0r}-\alpha_{0\ell} - 1]_q }
\end{align*}
where we have used
\begin{align*}
a_{0r}-q^{-2}a_{0\ell}-q^{-1} &= q^{-2}(q^2a_{0r}-a_{0\ell}-q)\\
&= q^{-(\alpha_{0r}+\alpha_{0\ell}+1)}[\alpha_{0r}-\alpha_{0\ell}-1]_q,\\
a_k - a_{0r} &= q^{-(\alpha_{0r}+\alpha_k)}[\alpha_k - \alpha_{0r}]_q.
\end{align*}
Therefore we obtain
$$
\mu_r = (-1)^{n-1}q^{\eta_r}\frac{ \prod_{k=1}^n[\alpha_k - \alpha_{0r}]_q }{
\prod_{\ell\neq r}^{n-1} [\alpha_{0r}-\alpha_{0\ell} - 1]_q }
$$
where 
\begin{align*}
\eta_r &= \sum_{\ell\neq r}^{n-1} (\alpha_{0r}+\alpha_{0\ell}+1)-\sum_{k=1}^n(\alpha_{0r} +
\alpha_k)
\\
&= -3\alpha_{0r} + 1+\sum_{\ell=1}^{n-1}\Lambda_{0\ell} - \sum_{k=1}^n\Lambda_k
\\
&= -3\alpha_{0r}+1 - \theta(\Lambda_0,\Lambda).
\end{align*}
Utilising the above formulae for $\omega_k$ and $\mu_r$ we have for the invariants
\begin{align*}
\omega_{kr} &= (a_k-a_{0r})^{-1} (a_k - q^{-2}a_{0r}-q^{-1})^{-1}\omega_k\mu_r
\\
&= (-1)^{n-1}  
\frac{ \prod_{\ell\neq r}^{n-1} (a_k - q^{-2}a_{0\ell}-q^{-1}) }{ \prod_{m\neq k}^n(a_k-a_m) }  
\frac{ \prod_{p\neq k}^n(a_p-a_{0r}) }{ \prod_{s\neq r}^{n-1} (a_{0r} - q^{-2}a_{0s} - q^{-1}) }
\\
&= (-1)^{n-1}\prod_{\ell\neq r}^{n-1}
\left( \frac{a_k - q^{-2}a_{0\ell} - q^{-1}}{a_{0r} -q^{-2}a_{0\ell} - q^{-1}} \right)
\prod_{p\neq k}^n
\left( \frac{a_p - a_{0r}}{a_k - a_p} \right)
\\
&= \prod_{\ell\neq r}^{n-1}
\left( \frac{a_k - q^{-2}a_{0\ell} - q^{-1}}{a_{0r} -\ q^{-2}a_{0\ell} - q^{-1}} \right)
\prod_{p\neq k}^n
\left( \frac{a_p - a_{0r}}{a_p - a_k} \right)
\\
&= \prod_{\ell\neq r}^{n-1}
\frac{ q^{-(\alpha_k + \alpha_{0\ell}+1)}[\alpha_k-\alpha_{0\ell}-1]_q }{
q^{-(\alpha_{0r}+\alpha_{0\ell}+1)}[\alpha_{0r}-\alpha_{0\ell} - 1]_q }
\prod_{p\neq k}^n 
\frac{ q^{-(\alpha_p + \alpha_{0r})}[\alpha_p - \alpha_{0r}]_q }{ q^{-(\alpha_p+\alpha_k)}[\alpha_p
- \alpha_k]_q }
\\
&= q^{\alpha_k - \alpha_{0r}}
\prod_{\ell\neq r}^{n-1} \frac{ [\alpha_k - \alpha_{0\ell} - 1]_q }{ [\alpha_{0r} - \alpha_{0\ell} -
1]_q }
\prod_{p\neq k}^n\frac{ [\alpha_p-\alpha_{0r}]_q }{ [\alpha_p-\alpha_k]_q }.
\end{align*}

Similarly, 
\begin{align*}
\tilde{\omega}_{kr} &=
( \tilde{a}_k - \tilde{a}_{0r} )^{-1}
( \tilde{a}_k - q^2 \tilde{a}_{0r}+q )^{-1}
\tilde{\omega}_k\tilde{\mu}_r
\\
&= (-1)^{n-1}
\frac{ \prod_{\ell\neq r}^{n-1}(\tilde{a}_k-q^2\tilde{a}_{0\ell}+1) }{ \prod_{\ell\neq k}^n
(\tilde{a}_k-\tilde{a}_\ell) }
\frac{ \prod_{p\neq k}^n(\tilde{a}_p-\tilde{a}_{0r}) }{ \prod_{\ell\neq r}^{n-1}(\tilde{a}_{0r} -
q^2\tilde{a}_{0\ell}+q) } \\
&= \prod_{\ell\neq r}^{n-1}\left( \frac{ \tilde{a}_k-q^2\tilde{a}_{0\ell}+q }{ \tilde{a}_{0r} -
q^2\tilde{a}_{0\ell}+q } \right)
\prod_{p\neq k}^k\left( \frac{ \tilde{a}_p - \tilde{a}_{0r} }{ \tilde{a}_p-\tilde{a}_k } \right) \\
&= \prod_{\ell\neq r}^{n-1} 
\frac{
q^{1-\overline{\alpha}_k-\overline{\alpha}_{0\ell}}[\overline{\alpha}_k-\overline{\alpha}_{0\ell} +
1]_q }{
q^{1-\overline{\alpha}_{0r}-\overline{\alpha}_{0\ell}}[\overline{\alpha}_{0r}-\overline{\alpha}_{0\ell}+1]_q }
\prod_{p\neq k}^n
\frac{
q^{-(\overline{\alpha}_p+\overline{\alpha}_{0r})}[\overline{\alpha}_p-\overline{\alpha}_{0r}]_q }{
q^{-(\overline{\alpha}_p+\overline{\alpha}_k)}[\overline{\alpha}_p-\overline{\alpha}_k]_q } \\
&= q^{\overline{\alpha}_k-\overline{\alpha}_{0r}}
\prod_{\ell\neq r}^{n-1}
\frac{ [\overline{\alpha}_k-\overline{\alpha}_{0\ell}+1]_q }{
[\overline{\alpha}_{0r}-\overline{\alpha}_{0\ell}+1]_q }
\prod_{p\neq k}^n 
\frac{ [\overline{\alpha}_p - \overline{\alpha}_{0r}]_q }{ [\overline{\alpha}_p -
\overline{\alpha}_k]_q }.
\end{align*}


\section{On matrix elements} \label{Section6}

\subsection{Matrix elements of elementary operators $A_{m\ m+1}$, $A_{m+1\ m}$, $\tilde{A}_{m\
m+1}$, $\tilde{A}_{m+1\ m}$}

Here we denote the characteristic roots of the canonical subalgebra $U_q(m)$ by
$$
\tilde{a}_{k,m},\ \ a_{k,m}
$$
so that
$$
a_{k,m} = \frac{1-q^{-2\alpha_{k,m}}}{q-q^{-1}},\ \ \tilde{a}_{k,m} =
\frac{1-q^{-2\overline{\alpha}_{k,m}}}{q-q^{-1}}
$$
with $\alpha_{k,m} = \Lambda_{k,m} + m-k,$ $\overline{\alpha}_{k,m} = \Lambda_{k,m} + 1-k.$ We also
denote the $U_q(m)$ (dual) vector operators by
\begin{align*}
\tilde{\phi}_{m,i} &= \tilde{A}_{i\ m+1},\ \ 1\leq i\leq m,\\
\tilde{\psi}_{m,i} &= \tilde{A}_{m+1\ i},\ \ 1\leq i\leq m,
\end{align*}
and we denote the $U_q(m)$ analogues of the $U_q(n-1)$ invariants $\omega_k,$ $\tilde{\omega}_k,$
$\mu_r$, $\tilde{\mu}_r$ by $\omega_{k,m},$ $\tilde{\omega}_{k,m},$ $\mu_{r,m},$ $\tilde{\mu}_{r,m}$
respectively. Thus with this notation we have
$$
\omega_k\equiv \omega_{k,n},\ \ \tilde{\omega}_k\equiv \tilde{\omega}_{k,n},\ \ \mu_r\equiv
\mu_{r,n-1},\ \ \tilde{\mu}_r\equiv \tilde{\mu}_{r,n-1}. 
$$

Now let $(\Lambda)$ be a Gelfand-Tsetlin pattern whose $m$th row is given by
the $U_q(m)$ highest weight ($\Lambda_{1m},\Lambda_{2m},\cdots,\Lambda_{mm})$.
Then the action of the elementary operator $\tilde{\psi}_{m,m}=\tilde{A}_{m+1\
m}$ on the corresponding Gelfand-Tsetlin state is given by
$$
\tilde{\psi}_{m,m}|(\Lambda)\rangle =
\sum_{r=1}^m\tilde{\psi}_m[r]_m|(\Lambda)\rangle
$$
where
$$
\tilde{\psi}_m[r]_m|(\Lambda)\rangle =
\tilde{M}_{r,m}|(\Lambda+\varepsilon_{r,m})\rangle.
$$
Choosing the positive real phases for these matrix elements, in agreement with
the classical convention, we have
\begin{align*}
	\tilde{M}_{r,m} &= \langle
	(\Lambda)|\tilde{\psi}_m[r]_m^\dagger\tilde{\psi}_m[r]_m|(\Lambda)\rangle^{1/2}\\
	&=
	\langle (\Lambda)|\tilde{\phi}_m[r]_m\tilde{\psi}_m[r]_m|(\Lambda)\rangle^{1/2}\\
	&=
	\langle (\Lambda)|\tilde{\mu}_{r,m}\tilde{\omega}_{r,m}|(\Lambda)\rangle^{1/2},
\end{align*}
where, in the notation above,
$$
\tilde{\mu}_{r,m}\tilde{\omega}_{r,m} = 
\frac{ (-1)^m \prod_{k=1}^{m+1}(\tilde{a}_{k,m+1} - \tilde{a}_{r,m})
\prod_{\ell=1}^{m-1}(\tilde{a}_{r,m} - q^2\tilde{a}_{\ell,m-1}+q) }
{ \prod_{\ell\neq r}^{m}(\tilde{a}_{r,m} -
q^2\tilde{a}_{\ell,m}+q)(\tilde{a}_{r,m} - \tilde{a}_{\ell,m}) }.
$$
\underline{Note}: The above agrees with the expression for the elementary
generator matrix elements in the classical limit $q\rightarrow 1$.

Similarly we consider the (dual/pseudo) vector operators
$$
\overline{\phi}_{i,m} = A_{m+1\ i},\  \ \overline{\psi}_{i,m} = A_{i\ m+1},\ \
1\leq i\leq m.
$$
Then the action of $\overline{\phi}_{m,m} = A_{m+1\ m}$ on a Gelfand-Tsetlin
state is given by 
$$
A_{m+1\ m} |(\Lambda)\rangle =
\sum_{r=1}^m\overline{\phi}_m[r]_m|(\Lambda)\rangle,
$$
where
$$
\overline{\phi}_m[r]_m|(\Lambda)\rangle =
\overline{M}_{r,m}|(\Lambda-\varepsilon_{r,m})\rangle.
$$
Again, using positive real phases, we obtain
\begin{align*}
	\overline{M}_{r,m} &=
	\langle(\Lambda)|\overline{\phi}_m[r]_m^\dagger\overline{\phi}_m[r]_m|(\Lambda)\rangle^{1/2}\\
	&= 
	\langle(\Lambda)|\overline{\psi}_m[r]_m\overline{\phi}_m[r]_m|(\Lambda)\rangle^{1/2}\\
	&= 
	\langle(\Lambda)|\mu_{r,m}\omega_{r,m}|(\Lambda)\rangle^{1/2}
\end{align*}
where
$$
\mu_{r,m}\omega_{r,m} =
\frac{ (-1)^m \prod_{k=1}^{m+1}({a}_{k,m+1} - {a}_{r,m})
\prod_{\ell=1}^{m-1}({a}_{r,m} - q^{-2}{a}_{\ell,m-1}-q^{-1}) }
{ \prod_{\ell\neq r}^{m}({a}_{r,m} -
q^{-2}{a}_{\ell,m}-q^{-1})({a}_{r,m} - {a}_{\ell,m}) }.
$$
Thus we obtain the following matrix element formulae
$$
\tilde{M}_{r,m} = q^{\frac12(\tilde{\eta}_{r,m}+\tilde{\xi}_{r,m})}\left\{
	\frac{\prod_{k=1}^{m+1}[\overline{\alpha}_{k,m+1} -
	\overline{\alpha}_{r,m}]_q\prod_{\ell=1}^{m-1}[\overline{\alpha}_{r,m}
	- \overline{\alpha}_{\ell,m-1}+1]_q}{\prod_{\ell\neq
	r}^m[\overline{\alpha}_{r,m}-\overline{\alpha}_{\ell,m}+1]_q[\overline{\alpha}_{r,m}-\overline{\alpha}_{\ell,m}]_q}
\right\}^{1/2},
$$
where, using notation extended from (\ref{xitilde}) and (\ref{etatilde}),
\begin{align*}
\tilde{\eta}_{r,m}+\tilde{\xi}_{r,m} 
&=
\sum_{j=1}^m\Lambda_{j,m} - \sum_{\ell=1}^{m-1}\Lambda_{\ell,m-1} -
\overline{\alpha}_{r,m} - 3\overline{\alpha}_{r,m}+1 -
\sum_{k=1}^{m+1}\Lambda_{k,m+1} + \sum_{j=1}^m\Lambda_{j,m}\\
	&= 1-4\overline{\alpha}_{r,m} + 2\sum_{j=1}^m\Lambda_{j,m} -
	\sum_{\ell=1}^{m-1}\Lambda_{\ell,m-1} -
	\sum_{k=1}^{m+1}\Lambda_{k,m+1}\\
	&= 1-4\overline{\alpha}_{r,m}+(\nu,\varepsilon_m-\varepsilon_{m+1})
\end{align*}
where $\nu$ is the weight of the (initial) Gelfand-Tsetlin state
$|(\Lambda)\rangle$
\begin{equation}
\Rightarrow \ \ \tilde{M}_{r,m} =
	q^{\frac12(\nu+\rho,\varepsilon_m-\varepsilon_{m+1}) -
2\overline{\alpha}_{r,m}}\left\{
	\frac{\prod_{k=1}^{m+1}[\overline{\alpha}_{k,m+1} -
	\overline{\alpha}_{r,m}]_q\prod_{\ell=1}^{m-1}[\overline{\alpha}_{r,m}
	- \overline{\alpha}_{\ell,m-1}+1]_q}{\prod_{\ell\neq
	r}^m[\overline{\alpha}_{r,m}-\overline{\alpha}_{\ell,m}+1]_q[\overline{\alpha}_{r,m}-\overline{\alpha}_{\ell,m}]_q}
\right\}^{1/2}.
\label{p57star}
\end{equation}
Similarly we obtain
\begin{equation}
	\overline{M}_{r,m} =
	q^{\frac12(\nu+\rho,\varepsilon_m-\varepsilon_{m+1}) -
2{\alpha}_{r,m}}\left\{
	\frac{\prod_{k=1}^{m+1}[{\alpha}_{k,m+1} -
	{\alpha}_{r,m}]_q\prod_{\ell=1}^{m-1}[{\alpha}_{r,m}
	- {\alpha}_{\ell,m-1}-1]_q}{\prod_{\ell\neq
	r}^m[{\alpha}_{r,m}-{\alpha}_{\ell,m}-1]_q[{\alpha}_{r,m}-{\alpha}_{\ell,m}]_q}
\right\}^{1/2}.
\label{p57starstar}
\end{equation}
\underline{Note}: Here, 
$\rho$ is the $U_q(n)$ Weyl vector, so
$(\rho,\varepsilon_m - \varepsilon_{m+1}) = 1$.

For the remaining elementary generators $\tilde{A}_{m\ m+1}$, $A_{m\ m+1}$ we
have, from Appendix E,
$$
\tilde{A}_{m\ m+1} = \tilde{\phi}_{m,m} = (\tilde{A}_{m+1\ m})^\dagger =
\tilde{\psi}_{m,m}^\dagger
$$
so that on a Gelfand-Tsetlin state $|(\Lambda)\rangle$ we have
$$
\tilde{A}_{m\ m+1}|(\Lambda)\rangle =
\sum_{r=1}^m\tilde{\phi}_m[r]_m|(\Lambda)\rangle,
$$
where
$$
\tilde{\phi}_m[r]_m|(\Lambda)\rangle = \tilde{M}_{r,m}' |(\Lambda -
\varepsilon_{r,m})\rangle,
$$
and
$$
\tilde{M}_{r,m}' = \langle(\Lambda - \varepsilon_{r,m})|\tilde{A}_{m\
m+1}|(\Lambda)\rangle = \langle(\Lambda)|\tilde{A}_{m+1\ m}|(\Lambda -
\varepsilon_{r,m})\rangle.
$$
From the above formulae for the matrix elements $\tilde{M}_{r,m}$ of
$\tilde{A}_{m+1\ m}$, we obtain
$$
\tilde{M}_{r,m}' = q^{\frac12(\nu+\rho,\varepsilon_m-\varepsilon_{m+1}) -
2\overline{\alpha}_{r,m} + 1}\left\{
	\frac{\prod_{k=1}^{m+1}[\overline{\alpha}_{k,m+1} -
	\overline{\alpha}_{r,m}+ 1]_q\prod_{\ell=1}^{m-1}[\overline{\alpha}_{r,m}
	- \overline{\alpha}_{\ell,m-1}]_q}{\prod_{\ell\neq
	r}^m[\overline{\alpha}_{r,m}-\overline{\alpha}_{\ell,m}]_q[\overline{\alpha}_{r,m}-\overline{\alpha}_{\ell,m}-1]_q}
\right\}^{1/2}.
$$
\underline{Note}: If $\nu$ is the weight of the initial Gelfand-Tsetlin state
$|(\Lambda)\rangle$, then the weight of $|(\Lambda-\varepsilon_{r,m})\rangle$
is actually $\nu-\varepsilon_m+\varepsilon_{m+1}$.

Similarly we have for the generators $A_{m\  m+1}$,
$$
A_{m\ m+1}|(\Lambda)\rangle = \overline{\psi}_{m,m}|(\Lambda)\rangle =
\sum_{r=1}^m\overline{\psi}_m[r]_m|(\Lambda)\rangle,
$$
where
$$
\overline{\psi}_m[r]_m|(\Lambda)\rangle = M_{r,m}'|(\Lambda +
\varepsilon_{r,m})\rangle.
$$
We obtain
\begin{align*}
	M_{r,m}' &= \langle(\Lambda+\varepsilon_{r,m})|A_{m\
	m+1}|(\Lambda)\rangle \\
	&= \langle(\Lambda)| A_{m+1\ m}|(\Lambda+\varepsilon_{r,m})\rangle\\
	&= \overline{M}_{r,m}(\Lambda+\varepsilon_{r,m})\\
	&= q^{\frac12(\nu+\rho,\varepsilon_m-\varepsilon_{m+1}) -
2{\alpha}_{r,m}+1}\left\{
	\frac{\prod_{k=1}^{m+1}[{\alpha}_{k,m+1} -
	{\alpha}_{r,m}-1]_q\prod_{\ell=1}^{m-1}[{\alpha}_{r,m}
	- {\alpha}_{\ell,m-1}]_q}{\prod_{\ell\neq
	r}^m[{\alpha}_{r,m}-{\alpha}_{\ell,m}]_q[{\alpha}_{r,m}-{\alpha}_{\ell,m}+1]_q}
\right\}^{1/2}.
\end{align*}


\subsection{Matrix elements of quantum group generators}

From the transformation properties (\ref{p18star}) (set $k=\ell=m+1,$ $i=m$) we
have
\begin{align*}
	[e_m,\tilde{A}_{m+1\  m+1}] &=
	q^{-\frac12(\varepsilon_m-\varepsilon_{m+1},\varepsilon_{m+1})}\tilde{A}_{m+1\
	m}q^{-\frac12h_m}\\
	&= q^{\frac12}\tilde{A}_{m+1\ m} q^{-\frac12h_m}.
\end{align*}
On the other hand, note that the first order invariant
$\tilde{C}_1=\tilde{C}_{1,m+1}$ for
$U_q(m+1)$ is expressible
$$
\tilde{C}_{1,m+1} = q\tilde{C}_{1,m} + q^{-m}\tilde{A}_{m+1\ m+1}\  \
\Rightarrow\ \ \tilde{A}_{m+1\ m+1} = q^m[\tilde{C}_{1,m+1} -
q\tilde{C}_{1,m}],
$$
where $\tilde{C}_{1,m}$ denotes the first order invariant of $U_q(m)$.
Since $\tilde{C}_{1,m+1}$ commutes with $e_m$ this gives
$$
q^{m+1}[\tilde{C}_{1,m},e_m] = q^{\frac12}\tilde{A}_{m+1\ m}q^{-\frac12h_m}
$$
or
$$
[\tilde{C}_{1,m},e_m] = q^{-(m+\frac12)}\tilde{A}_{m+1\ m}q^{-\frac12 h_m}.
$$
It follows that the non-zero matrix elements of $e_m$ are of the form
$$
N_{r,m} = \langle(\Lambda+\varepsilon_{r,m})|e_m|(\Lambda)\rangle
$$
and are related to the matrix elements
$$
\tilde{M}_{r,m} = \langle(\Lambda+\varepsilon_{r,m})|\tilde{A}_{m+1\
m}|(\Lambda)\rangle
$$
by
\begin{align*}
	q^{-(m+\frac12)-\frac12(\nu,\varepsilon_m-\varepsilon_{m+1})}\tilde{M}_{r,m} &=
	\langle(\Lambda+\varepsilon_{r,m})|\tilde{C}_{1,m}e_m - e_m
	\tilde{C}_{1,m}|(\Lambda)\rangle\\
	&= [\tilde{C}_{1,m}(\Lambda+\varepsilon_{r,m}) -
	\tilde{C}_{1,m}(\Lambda)]N_{r,m}.
\end{align*}
Now from equation (\ref{p35star}) we have
\begin{align*}
	\tilde{C}_{1,m}(\Lambda+\varepsilon_{r,m}) - \tilde{C}_{1,m}(\Lambda) &=
	\sum_{i=1}^m
	\left( q^{-(\Lambda+2\rho+\varepsilon_r,\varepsilon_i)}[(\Lambda+\varepsilon_r,\varepsilon_i)]_q
	- q^{-(\Lambda+2\rho,\varepsilon_i)}[(\Lambda,\varepsilon_i)]_q\right)
	\\
	&=
	q^{-(\Lambda+2\rho,\varepsilon_r)-1}[(\Lambda+\varepsilon_r,\varepsilon_r)]_q
	- q^{-(\Lambda+2\rho,\varepsilon_r)}[(\Lambda,\varepsilon_r)]_q\\
	&= q^{-2(\Lambda+\rho,\varepsilon_r)-1} = q^{-2\overline{\alpha}_{r,m} -
	m}
\end{align*}
$$
\Rightarrow \ \ q^{-(m+\frac12) - \frac12(\nu,\varepsilon_m-\varepsilon_{m+1})}\tilde{M}_{r,m} =
q^{-2\overline{\alpha}_{r,m} - m}N_{r,m}
$$
or
$$
N_{r,m} = q^{2\overline{\alpha}_{r,m} -
\frac12(\nu+\rho,\varepsilon_m-\varepsilon_{m+1})}\tilde{M}_{r,m}
$$
where,  again, $\nu$ is the weight of the (initial) Gelfand-Tsetlin state
$|(\Lambda)\rangle$. By comparison with equation (\ref{p57star}) this gives
$$
N_{r,m} = \left\{ \frac{\prod_{k=1}^{m+1}[\overline{\alpha}_{k,m+1} -
	\overline{\alpha}_{r,m}]_q\prod_{\ell=1}^{m-1}[\overline{\alpha}_{r,m}
	- \overline{\alpha}_{\ell,m-1}+1]_q}{\prod_{\ell\neq
	r}^m[\overline{\alpha}_{r,m}-\overline{\alpha}_{\ell,m}+1]_q[\overline{\alpha}_{r,m}-\overline{\alpha}_{\ell,m}]_q}
	\right\}^{1/2}.
$$
Similarly from equation (\ref{p18star}) with $i=m,$ $k=\ell=m+1$ we obtain
\begin{align*}
[f_m,\tilde{A}_{m+1\ m+1}] & = -q^{3/2}\tilde{A}_{m\ m+1}q^{-h_m/2}\\
	\Rightarrow \ \  -q^{3/2}\tilde{A}_{m\ m+1}q^{-h_m/2} & =
	[f_m,\tilde{A}_{m+1\ m+1}]\\
	& =-q^{m+1}[f_m,\tilde{C}_{1,m}],
\end{align*}
or
$$
[\tilde{C}_{1,m},f_m] = -q^{-m+\frac12}\tilde{A}_{m\ m+1}q^{-h_m/2}.
$$
Therefore the non-zero matrix elements of $f_m$ are of the form 
$$
\overline{N}_{r,m} = \langle (\Lambda - \varepsilon_{r,m})|f_m|(\Lambda) \rangle
$$
and are related to the matrix elements
$$
\tilde{M}'_{r,m} = \langle (\Lambda-\varepsilon_{r,m})|\tilde{A}_{m\
m+1}|(\Lambda) \rangle
$$
by
$$
[\tilde{C}_{1,m}(\Lambda-\varepsilon_{r,m})-\tilde{C}_{1,m}(\Lambda)]\overline{N}_{r,m}
= -q^{-m+\frac12-\frac12(\varepsilon_m-\varepsilon_{m+1},\nu)}\tilde{M}'_{r,m}
$$
where $\nu$ is the weight of the Gelfand-Tsetlin state $|(\Lambda)\rangle$.

Now we note that 
\begin{align*}
\tilde{C}_{1,m}(\Lambda-\varepsilon_{r,m})-\tilde{C}_{1,m}(\Lambda) 
&= q^{-(\Lambda+2\rho-\varepsilon_r,\varepsilon_r)}
	[(\Lambda-\varepsilon_r,\varepsilon_r)]_q -
	q^{-(\Lambda+2\rho,\varepsilon_r)}[(\Lambda,\varepsilon_r)]_q \\
&= -q^{-2(\Lambda+\rho,\varepsilon_r)+1} = -q^{-m-2\overline{\alpha}_{r,m} +
	2}
\end{align*}
\begin{align*}
\Rightarrow \ \ \overline{N}_{r,m} &= q^{2\overline{\alpha}_{r,m}
	-\frac32-\frac12(\nu,\varepsilon_m-\varepsilon_{m+1})}\tilde{M}'_{r,m}\\
& = \left\{
	\frac{\prod_{k=1}^{m+1}[\overline{\alpha}_{k,m+1}-\overline{\alpha}_{r,m}+1]_q\prod_{\ell=1}^{m-1}[\overline{\alpha}_{r,m}
	- \overline{\alpha}_{\ell,m-1}]_q}{ \prod_{\ell\neq
	r}^m[\overline{\alpha}_{r,m} -
	\overline{\alpha}_{\ell,m}]_q[\overline{\alpha}_{r,m} -
	\overline{\alpha}_{\ell,m} - 1]_q } \right\}^{1/2}.
\end{align*}
\underline{Note}: In the above, $\nu$ denotes the weight of the initial
Gelfand-Tsetlin state $|(\Lambda)\rangle$ as before. Also recall that
$-\frac12 - \frac12(\nu,\varepsilon_m-\varepsilon_{m+1}) =
-\frac12(\nu+\rho,\varepsilon_m-\varepsilon_{m+1})$.

Finally,  in terms of characteristic roots, 
$$
\alpha_{k,m} = \overline{\alpha}_{k,m}+m-1 \ \ \Rightarrow \ \
\overline{\alpha}_{k,m} = \alpha_{k,m}-m+1,
$$
we obtain the following matrix element formulae for the simple generators:
\begin{align*}
N_{r,m} &= \left\{
	\frac{\prod_{k=1}^{m+1}[{\alpha}_{k,m+1}-{\alpha}_{r,m}-1]_q\prod_{\ell=1}^{m-1}[{\alpha}_{r,m}
	- {\alpha}_{\ell,m-1}]_q}{ \prod_{\ell\neq
	r}^m[{\alpha}_{r,m} -
	{\alpha}_{\ell,m}+1]_q[{\alpha}_{r,m} -
	{\alpha}_{\ell,m} ]_q } \right\}^{1/2},
\end{align*}
\begin{align*}
\overline{N}_{r,m} &= \left\{
	\frac{\prod_{k=1}^{m+1}[{\alpha}_{k,m+1}-{\alpha}_{r,m}]_q\prod_{\ell=1}^{m-1}[{\alpha}_{r,m}
	- {\alpha}_{\ell,m-1}-1]_q}{ \prod_{\ell\neq r}^m
	[{\alpha}_{r,m} - {\alpha}_{\ell,m}]_q
	[{\alpha}_{r,m} - {\alpha}_{\ell,m} - 1]_q } \right\}^{1/2}.
\end{align*}
These expressions agree with the previous results of Gould et al.
\cite{GLB1992,Gould1992}.


\section*{Appendix A: vector and contragredient vector operators}

Following equation (\ref{doublestarp6}) we call a collection of components
$\psi\equiv \{ \psi_i\}_{i=1}^n$ a {\em vector operator} if the following
transformation law holds:
$$
a\psi_i = \pi_0(a_{(1)})_{ji}\psi_j a_{(2)},\ \ a\in U_q(n),
$$
where $\pi_0$ is the defining (vector) representation (which is undeformed). In
terms of quantum group generators, this transformation law is expressible
\begin{align*}
q^{E_{kk}} \psi_i &= q^{\delta_{ki}}\psi_iq^{E_{kk}},\\
e_k\psi_i &= \pi_0(q^{h_k/2})_{ji}\psi_j e_k + \pi_0(E_{k\
	k+1})_{ji}\psi_jq^{-h_k/2}\\
	&= q^{\frac12(\varepsilon_k-\varepsilon_{k+1},\varepsilon_i)}\psi_ie_k + \delta_{i\ k+1}\psi_k
	q^{-h_k/2}\\
	f_k\psi_i &= \pi_0(q^{h_k/2})_{ji}\psi_j f_k + \pi_0(E_{k+1\ k})_{ji}
	\psi_j q^{-h_k/2}\\
	&= q^{\frac12(\varepsilon_k-\varepsilon_{k+1},\varepsilon_i)}\psi_i f_k + \delta_{ik}
	\psi_{k+1} q^{-h_k/2}.
\end{align*}
In terms of the $q$-bracket
$$
[x_k,\psi_i]_{q_i} = x_k \psi_i - q^{\frac12(\varepsilon_k-\varepsilon_{k+1},\varepsilon_i)} \psi_i
x_k,\ \ x_k = e_k,f_k,
$$
the last two relations are expressible
\begin{align*}
	[e_k,\psi_i]_{q_i} &= \delta_{i \ k+1} \psi_k q^{-h_k/2}\\
	[f_k,\psi_i]_{q_i} &= \delta_{ik} \psi_{k+1} q^{-h_k/2}.
\end{align*}

Dually, we call a collection of operators $\phi\equiv \{\phi_i\}_{i=1}^n$ a
{\em dual vector operator} if the following transformation law holds:
\begin{equation}
	a\phi_i = \pi_0^*(a_{(1)})_{ji}\phi_j a_{(2)},\ a\in U_q(n),
	\label{pA2star}
\end{equation}
where $\pi_0^*$ is the dual vector representation defined by 
$$
\pi_0^*(a)_{ji} = \langle e_j,ae_i^*\rangle = \langle S(a)e_j,e_i^*\rangle =
\pi_0(S(a))_{ij}.
$$
\underline{Note}: $\overline{\phi}_i \equiv q^{(\rho,\varepsilon_i)}\phi_i$
transforms as $\overline{\pi}_0$, called a {\em pseudo vector operator}. Thus
$$
a\overline{\phi}_i = \overline{\pi}_0(a_{(1)})_{ji}\overline{\phi}_ja_{(2)},\ \
a\in U_q(n).
$$
In terms of elementary generators, the above transformation law (\ref{pA2star})
is expressible
\begin{align*}
q^{E_{kk}} \phi_i &= q^{-\delta_{ki}}\phi_iq^{E_{kk}},\\
	e_k\phi_i &= \pi_0^*(q^{h_k/2})_{ji}\phi_j e_k -q^{-1} \pi_0(E_{k\
	k+1})_{ij}\phi_jq^{-h_k/2}\\
\Rightarrow \ \ e_k\phi_i - q^{-\frac12(\varepsilon_k-\varepsilon_{k+1},\varepsilon_i)}\phi_ie_k &= -
	q^{-1} \delta_{ki}\phi_{k+1}q^{-h_k/2}
\end{align*}
and similarly
$$
f_k\phi_i - q^{-\frac12(\varepsilon_k-\varepsilon_{k+1},\varepsilon_i)}\phi_if_k = 
-q \delta_{i\ k+1} \phi_k q^{-h_k/2}.
$$

It is also worth noting that a collection of components $\overline{\psi}_i$
transforms as a tensor operator of rank $\overline{\pi}^*$ (dual pseudo vector
operator) if and only if it satisfies
$$
a\overline{\psi}_i = \overline{\pi}_0^*(a_{(1)})_{ji}\overline{\psi}_j a_{(2)}.
$$
In terms of elementary generators we have
$$
e_k\overline{\psi}_i = q^{\frac12(\varepsilon_k-\varepsilon_{k+1},\varepsilon_i)}\overline{\psi}_ie_k
+ q^{(\rho,\varepsilon_i-\varepsilon_j)}(E_{k\ k+1})_{ji}\overline{\psi}_j
q^{-h_k/2}
$$
$$
\Rightarrow \ \ e_k\overline{\psi}_i -
q^{\frac12(\varepsilon_k-\varepsilon_{k+1},\varepsilon_i)}\overline{\psi}_ie_k = \delta_{i\
k+1}q^{-1}\overline{\psi}_kq^{-h_k/2},
$$
and similarly
\begin{align*}
	f_k \overline{\psi}_i - q^{\frac12(\varepsilon_k-\varepsilon_{k+1},\varepsilon_i)}\overline{\psi}_if_k &= 
	q^{(\rho,\varepsilon_i-\varepsilon_j)}(E_{k+1\ k})_{ji}\overline{\psi}_j
	q^{-h_k/2}\\
	&= \delta_{ik} q\overline{\psi}_{k+1} q^{-h_k/2}
\end{align*}
with 
$$
q^{E_{kk}}\overline{\psi}_i = q^{(\varepsilon_k,\varepsilon_i)}
\overline{\psi}_i q^{E_{kk}}.
$$
\underline{Note}: If $\psi_i$ is a vector operator then
$$
\overline{\psi}_i = q^{(\rho,\varepsilon_i)}\psi_i
$$
transforms as $\overline{\pi}^*$ and conversely, i.e. if $\overline{\psi}_i$
transforms as $\overline{\pi}^*$ then
$$
\psi_i = q^{-(\rho,\varepsilon_i)}\overline{\psi}_i
$$
transforms as a vector operator.


\section*{Appendix B: Shift components}

Let $\{\phi_i\}$ be a dual vector operator of $U_q(n)$ which, acting on an
irreducible module $V(\Lambda)$, determines an intertwining operator $\phi$:
$$
\phi(e_i\otimes v) = \phi_i v,\ \ v\in V(\Lambda).
$$
As for vector operators, $\phi$ may be resolved into shift components
$$
\phi_i = \sum_{r=1}^n\phi[r]_i
$$
where
$$
\phi[r](V^*\otimes V(\Lambda)) \cong V(\Lambda - \varepsilon_r).
$$
Here we show that the projection operators $\tilde{P}_r$ of equation
(\ref{p24star}) project out the above shift components from the left. We first
need (c.f. Proposition \ref{Prop2})

\begin{lemma}
\label{lemmaB}
The identity module occurs exactly once in $V\otimes V^*$ and is spanned by the
	vector $\xi=e_\alpha\otimes e_\alpha^*$ (sum on $\alpha$), where
	$\{e_{\alpha}\}$ is a basis for $V$ with corresponding dual basis
	$\{e_\alpha^*\}$ for $V^*$ (regarded as a module under the usual action
	determined by the antipode $S$).
\end{lemma}
Proof:
\begin{align*}
	a\xi &= a_{(1)}e_\alpha\otimes a_{(2)}e_\alpha^*\\
	&= a_{(1)}e_\alpha\otimes\langle a_{(2)}e_\alpha^*,e_\beta \rangle
	e_\beta^*\\
	&= a_{(1)} e_{\alpha}\langle e_\alpha^*,S(a_{(2)})e_\beta\rangle
	\otimes e_\beta^*\\
	&= a_{(1)} S(a_{(2)})e_\beta\otimes e_\beta^* = \varepsilon(a)\xi.
\end{align*}
\begin{flushright}$\Box$\end{flushright}

Now let $\tilde{P}_r$ be the projection operators of equation
(\ref{p24star}) and $\phi[\ell]$ be the shift components of $\phi$. We
observe that
\begin{align*}
	(\id\otimes \phi)(\xi\otimes v) &= e_i\otimes \phi(e_i^*\otimes v),\ \
	v\in V(\Lambda)\\
	&= e_i\otimes \phi_iv
\end{align*}
gives an isomorphic copy of $V(\Lambda)$. Now
$$
\tilde{P}_r(\id\otimes\phi[\ell])(\xi\otimes v) = \tilde{P}_r(e_i\otimes
\phi[\ell](e_i^*\otimes v)) \subseteq
V(\Lambda+\varepsilon_r-\varepsilon_\ell)\cap V(\Lambda)
$$
$$
\Rightarrow \tilde{P}_r(\id\otimes \phi[\ell])(\xi\otimes V(\Lambda)) = (0), \
\ r\neq \ell.
$$
Therefore we have
\begin{align*}
	e_i\otimes\phi[r]_iv &= e_i\otimes \phi[r](e_i^*\otimes v)\\
	&= \tilde{P}_r(\id\otimes \phi)(\xi\otimes v)\\
	&= \tilde{P}_r(e_i\otimes \phi(e_i^*\otimes v))\\
	&= e_j\otimes (\tilde{P}_r)_{ji}\phi(e_i^*\otimes v))\\
	&= e_j\otimes (\tilde{P}_r)_{ji}\phi_iv\\
\ \  \Rightarrow \ \ \phi[r]_i &= (\tilde{P}_r)_{ij}\phi_j.
\end{align*}
Thus $\tilde{P}_r$ projects out shift components of dual vector operators from
the {\em left}. However, $\tilde{P}_r$ clearly projects out the shift
components of a vector operator $\psi = \{\psi_i\}$ from the {\em right} since
\begin{align*}
	\psi[r]_iv &= \psi[r](e_i\otimes v)\\
	&= \psi\tilde{P}_r(e_i\otimes v)\\
	&= \psi(e_j\otimes (\tilde{P}_r)_{ji}v)\\
	&= \psi_j(\tilde{P}_r)_{ji}v,
\end{align*}
i.e. 
$$\psi[r]_i = \psi_j(\tilde{P}_r)_{ji}.$$
However, to project out these shift components from the left we need
\begin{lemma}
\label{lemmaBprime}
The identity module occurs exactly once in $V_0^*\otimes V_0$ and is spanned by
$$
\eta = e_i^*\otimes q^{2h_\rho}e_i = q^{2(\rho,\varepsilon_i)}e_i^*\otimes e_i
$$
(sum on $i$).
\end{lemma}
Proof: 
\begin{align*}
a\eta &= a_{(1)}e_i^*\otimes a_{(2)}q^{2h_\rho}e_i\\
&= \langle a_{(1)}e_i^*,e_j \rangle e_j^* \otimes a_{(2)}q^{2h_\rho}e_i\\
&= e_j^*\otimes a_{(2)}q^{2h_\rho}e_i \langle e_i^*,S(a_{(1)})e_j \rangle \\
&= e_j^*\otimes a_{(2)}q^{2h_\rho}S(a_{(1)})e_j\\
&= e_j^*\otimes a_{(2)}S^{-1}(a_{(1)})q^{2h_\rho} e_j = \varepsilon(a) \eta.
\end{align*}
\begin{flushright}$\Box$\end{flushright} 
\underline{Note}: This result can be regarded as a particular case of
Lemma \ref{lemmaB} if we note that the transformed basis $q^{2h_\rho}e_\alpha$
gives a basis for the representation $\pi^{**}$.

Now let $P_r$ on $V^*\otimes V(\Lambda)$ be the projection operator arising
from the matrix $A$, so here we regard $V^*$ as a module under the action
defined by $\overline{\pi}_0$, i.e.
$$
ae_i^* = \overline{\pi}_0(a)_{ji}e_j^*,
$$
or
$$
\langle ae_i^*,e_j \rangle = \langle e_i^*,\gamma(a)e_j.
$$
Under this action the invariant of Lemma \ref{lemmaBprime} becomes
$$
\overline{\eta} = e_i^*\otimes q^{h_\rho}e_i = q^{(\rho,\varepsilon_i)}e_i^*\otimes e_i =
q^{-h_\rho}e_i^*\otimes e_i
$$
since
\begin{align*}
	a\overline{\eta} &= a_{(1)}e_i^*\otimes a_{(2)}q^{h_\rho}e_i\\
	&= \langle e_i^*,\gamma(a_{(1)})e_j \rangle e_j^*\otimes
	a_{(2)}q^{h_\rho}e_i \\
	&= e_j^*\otimes a_{(2)}q^{h_\rho}\gamma(a_{(1)})e_j\\
	&= e_j^*\otimes a_{(2)} S^{-1}(a_{(1)}) q^{h_\rho}e_j\\
	&= \varepsilon(a)\overline{\eta}.
\end{align*}

Now let $\psi = \{\psi_i\}$ be a vector operator acting on $V(\Lambda)$.
Following our derivation above we have for $v\in V(\Lambda)$
\begin{align*}
	q^{-h_\rho}e_i^*\otimes \psi[r]_iv &= q^{-h_\rho}e_i^*\otimes
	\psi[r](e_i\otimes v)\\
	&= (\id\otimes \psi[r])(\overline{\eta}\otimes v)\\
	&= P_r(\id\otimes \psi)(\overline{\eta}\otimes v)\\
	&= P_r(q^{-h_\rho}e_i^*\otimes \psi(e_i\otimes v))\\
	&= q^{(\rho,\varepsilon_i)}e_j^*\otimes (P_r)_{ji}\psi(e_i\otimes v)\\
	&= e_j^*\otimes (P_r)_{ji}q^{(\rho,\varepsilon_i)}\psi_iv\\
	&= e_i^*\otimes (P_r)_{ij} q^{(\rho,\varepsilon_j)}\psi_jv\\
\ \ \Rightarrow \ \ q^{(\rho,\varepsilon_i)}\psi[r]_i &=
	(P_r)_{ij}q^{(\rho,\varepsilon_j)}\psi_j.
\end{align*}
\underline{Note}: $\overline{\psi}_i = q^{(\rho,\varepsilon_i)}$ defines a dual
pseudo vector operator, i.e. transforming as $\overline{\pi}_0^*$. Then we have
$$
\overline{\psi}[r]_i = (P_r)_{ij}\overline{\psi}_j,
$$
thus showing that $P_r$ projects out the shift components of such tensor
operators from the {\em left}.

On the other hand, if $\overline{\phi}_i$ is a pseudo vector operator, then for
$v\in V(\Lambda)$
\begin{align*}
	\overline{\phi}[r]_iv &= \overline{\phi}[r](e_i^*\otimes v)\\
	&= \overline{\phi}P_r(e_i^*\otimes v)\\
	&= \overline{\phi}(e_j^*\otimes (P_r)_{ji}v)\\
	&= \overline{\phi}_j(P_r)_{ji}v\\
	\ \ \Rightarrow \ \ \overline{\phi}[r]_i &=
	\overline{\phi}_j(P_r)_{ji},
\end{align*}
showing that $P_r$ projects out the shift components of $\overline{\phi}$ from
the {\em right}.

\subsection*{Characteristic root shifts}

Let $\psi_i$ be a vector (or dual pseudo vector) operator and $\phi_i$ a dual
vector (or pseudo vector) operator. Then the shift components affect the
following shifts on the classical roots:
\begin{align*}
	\alpha_r \phi[r]_i &= \phi[r]_i(\alpha_r-1),\\
	\overline{\alpha}_r\phi[r]_i &= \phi[r]_i(\overline{\alpha}_r-1),\\
	\alpha_r\psi[r]_i &= \psi[r]_i(\alpha_r+1),\\
	\overline{\alpha}\psi[r]_i &= \psi[r]_i(\overline{\alpha}_r+1).
\end{align*}
Using
\begin{align*}
	\frac{1-q^{-2\alpha_r-2}}{q-q^{-1}} &=
	q^{-2}\frac{1-q^{-2\alpha_r}}{q-q^{-1}} + \frac{1-q^{-2}}{q-q^{-1}} =
	q^{-2}a_r + q^{-1},\\
	\frac{1-q^{-2\alpha_r+2}}{q-q^{-1}} &=
	q^{2}\frac{1-q^{-2\alpha_r}}{q-q^{-1}} + \frac{1-q^{2}}{q-q^{-1}} =
	q^2a_r - q,
\end{align*}
(and similarly for $\overline{\alpha}_r$ and $\tilde{a}_r$) we have the
following shifts
\begin{align*}
	a_r\psi[r]_i &= \psi[r]_i(q^{-2}a_r + q^{-1}),\\
	\tilde{a}_r\psi[r]_i &= \psi[r]_i(q^{-2}\tilde{a}_r+ q^{-1}),\\
	a_r\phi[r]_i &= \phi[r]_i(q^2a_r - q),\\
	\tilde{a}_r\phi[r]_i &= \phi[r]_i (q^2\tilde{a}_r - q),
\end{align*}
or equivalently
\begin{align*}
	\psi[r]_ia_r &= (q^2a_r-q)\psi[r]_i,\\
	\psi[r]_i \tilde{a}_r &= (q^2\tilde{a}_r - q)\psi[r]_i,\\
	\phi[r]_i a_r &= (q^{-2} a_r + q^{-1})\phi[r]_i,\\
	\phi[r]_i\tilde{a}_r &= (q^{-2}\tilde{a}_r+q^{-1})\phi[r]_i.
\end{align*}


\section*{Appendix C}

Here we show that the operators $\hat{E}_{ij}$ of equation (\ref{p17star}) are
related to those of equation (\ref{p17triplestar}), i.e. $\tilde{E}_{ij}$,
via action of the antipode.

First we recall the following definitions, with $i\lessgtr k\lessgtr j$:
\begin{align*}
	E_{ij} &= E_{ik}E_{kj} - q^{-1}E_{kj}E_{ik},\\
	E_{ij}' &= E_{ik}'E_{kj}' - qE_{kj}'E_{ik}'.
\end{align*}
Now observe that
$$
S^{-1}(E_{i\ i+1}) = -qE_{i\ i+1} = -qE_{i\ i+1}'
$$
\begin{align*}
	\Rightarrow \ \ S^{-1}(E_{i\ i+2}) &= 
S^{-1}(E_{i\ i+1}E_{i+1\  i+2} - q^{-1}E_{i+1\ i+2}E_{i\ i+1})\\
	&= q^2E_{i+1\ i+2}' E_{i\ i+1}' - qE_{i\ i+1}'E_{i+1\ i+2}'\\
	&= -q(E_{i\ i+1}'E_{i+1\ i+2}' - q^{-1}E_{i+1\ i+2}'E_{i\ i+1}')\\
	&= -qE_{i i+2}'.
\end{align*}
More generally, it follows by induction that
$$
S^{-1}(E_{ij}) = -q E_{ij}',\  \ i<j.
$$

In the case $i>j$, we observe first that
$$
S^{-1}(E_{i+1\ i}) = -q^{-1}E_{i+1\ i} = -q^{-1} E_{i+1\ i}'.
$$
Then
\begin{align*}
	S^{-1} &= S^{-1}(E_{i+2\ i+1}E_{i+1\ i} - q^{-1}E_{i+1\ i}E_{i+2\
	i+1})\\
	&= q^{-2}E_{i+1\ i}'E_{i+2\ i+1}' - q^{-3}E_{i+2\ i+1}'E_{i+1\ i}'\\
	&= -q^{-3}(E_{i+2\ i+1}'E_{i+1\ i} - qE_{i+1\ i}' E_{i+2\ i+1}')\\
	&= -q^{-3}E_{i+2\ i}'.
\end{align*}
More generally we observe that
$$
S^{-1}(E_{ij}) = -q^{(2\rho,\varepsilon_i - \varepsilon_j)+1}E_{ij}',\ \ i<j,
$$
i.e.
$$
S^{-1}(E_{ij}) = -q\left\{ 
\begin{array}{rl}  
	E_{ij}' ,& i<j\\
	q^{(2\rho,\varepsilon_i - \varepsilon_j)}E_{ij}' ,& i>j.
\end{array}
\right.
$$
Thus 
\begin{align*}
S^{-1}(\hat{E}_{ij}) &\stackrel{(\ref{p17star})}{=}
\left\{
\begin{array}{rl}
	(q-q^{-1})q^{-\frac12(E_{ii}+E_{jj}+1)}S^{-1}(E_{ij}),& i\neq j\\
	q^{-E_{ii}},& i=j
\end{array}
\right.
\\ &= 
\left\{
\begin{array}{rl}
	-(q-q^{-1})q^{-\frac12(E_{ii}+E_{jj}-1)}E_{ij}',& i< j\\
	q^{-E_{ii}},& i=j\\
	-(q-q^{-1})q^{-\frac12(E_{ii}+E_{jj}-1)}q^{(2\rho,\varepsilon_i-\varepsilon_j)}E_{ij}',
	& i>j
\end{array}
\right.
\\ &= 
\left\{
\begin{array}{rl}
	\tilde{E}_{ij},& i\leq j\\
	q^{(2\rho,\varepsilon_i-\varepsilon_j)}\tilde{E}_{ij},& i>j.
\end{array}
\right.
\end{align*}
Thus from equation (\ref{p26star})  we obtain
\begin{align*}
	(\pi_0^*\otimes \id)R &= \sum_{i\leq j} e_{ij}\otimes \tilde{E}_{ij} \\
	(\pi_0^*\otimes \id)R^T &= \sum_{i\leq j}
	q^{(2\rho,\varepsilon_i-\varepsilon_j)} e_{ji}\otimes \tilde{E}_{ji}.
\end{align*}
As we have seen, the dual vector representation utilised in reference
\cite{GLB1992} is
here denoted $\overline{\pi}_0$ (the pseudo vector representation) and is
defined by
$$
\overline{\pi}_0(a) = \pi_0^t(\gamma(a)) = \pi_0^*(q^{h_\rho}aq^{-h_\rho}),\ \
a\in U_q(n),
$$
so (c.f. equation (\ref{p26star}))
$$
\overline{\pi}_0(a)_{ij} = q^{(\rho,\varepsilon_j - \varepsilon_i)}
\pi_0^*(a)_{ij}.
$$
This gives the following expression for the $L$-operators of equations
(\ref{p27stara}) and (\ref{p27starb}) in terms of the $\tilde{E}_{ij}$:
\begin{align*}
	(\overline{\pi}_0\otimes \id)R &= \sum_{i\leq
	j}q^{(\rho,\varepsilon_i-\varepsilon_j)}e_{ij}\otimes \tilde{E}_{ij},\\
	(\overline{\pi}_0\otimes\id)R^T &= \sum_{i\leq j}
	q^{(\rho,\varepsilon_j-\varepsilon_i)} e_{ji}\otimes \tilde{E}_{ji}.
\end{align*}
This provides an alternative description of the $L$-operators and agrees with
reference \cite{GLB1992}.


\section*{Appendix D: Characteristic roots}

We have
\begin{align*}
	\tilde{a}_k &= \frac{1-q^{-2\overline{\alpha}_k}}{q-q^{-1}},\ \
	\overline{\alpha}_k = \Lambda_k+1-k,\\
	a_k &= \frac{1-q^{-2\alpha_k}}{q-q^{-1}},\ \ \alpha_k - \Lambda_k+n-k =
	\overline{\alpha}_k + n-1.
\end{align*}
Therefore
\begin{align*}
	\tilde{a}_k - q^2\tilde{a}_{0r} &= 
	\frac{ 1-q^{-2\overline{\alpha}_k}-q^2(1-q^{-2\overline{\alpha}_{0r}} )
		}{ q-q^{-1} } \\
	&= \frac{ q^{2-2\overline{\alpha}_{0r}} - q^{-2\overline{\alpha}_k} }{ q-q^{-1} }
	+ \frac{ q(q^{-1}-q) }{ q-q^{-1} } \\
\Rightarrow \ \ \tilde{a}_k-q^2\tilde{a}_{0r}+q &= \frac{
	q^{2-2\overline{\alpha}_{0r}} - q^{-2\overline{\alpha}_k} }{ q-q^{-1}}\\
	&= \frac{ q^{2-2(\alpha_{0r}+2-n)} - q^{-2(\alpha_k+1-n)} }{ q-q^{-1} }
	\\
&= q^{-2(1-n)}\frac{ (q^{-2\alpha_{0r}} - q^{-2\alpha_k} ) }{ q-q^{-1} } \\
	&= q^{-2(1-n)}(a_k-a_{0r}).
\end{align*}
Therefore we have the following identity:
$$
\tilde{a}_k-q^2\tilde{a}_{0r}+q =  q^{2(n-1)}(a_k-a_{0r}).
$$
Also, 
$$
\tilde{a}_k - \tilde{a}_\ell = \frac{ q^{-2\overline{\alpha}_\ell} -
q^{-2\overline{\alpha}_k} }{ q-q^{-1} }
= q^{-(\overline{\alpha}_\ell + \overline{\alpha}_k)}
\frac{ q^{\overline{\alpha}_k-\overline{\alpha}_\ell} -
q^{\overline{\alpha}_\ell-\overline{\alpha}_k} }{ q-q^{-1} }
$$
$$
\Rightarrow \ \ \tilde{a}_k - \tilde{a}_\ell = q^{-(\overline{\alpha}_\ell +
\overline{\alpha}_k)} [\overline{\alpha}_k - \overline{\alpha}_\ell]_q
$$
and similarly
\begin{align*}
	a_k - a_\ell &= \frac{q^{-2\alpha_\ell}-q^{-2\alpha_k} }{ q-q^{-1} } =
	q^{-(\alpha_\ell+\alpha_k)}[\alpha_k - \alpha_\ell]_q,\\
	a_k - a_{0r} &= \frac{q^{-2\alpha_{0r}} - q^{-2\alpha_k}}{q-q^{-1}} = 
	q^{-(\alpha_{0r}+\alpha_k)}[\alpha_k - \alpha_{0r}]_q,
\end{align*}
and so on. Also observe that
$$
\alpha_k - \alpha_\ell = \overline{\alpha}_k - \overline{\alpha}_\ell
$$
so that
\begin{align*}
	\tilde{a}_k - \tilde{a}_\ell &= q^{-(\overline{\alpha}_\ell +
	\overline{\alpha}_k)}[\alpha_k - \alpha_\ell]_q\\
	&= q^{2(n-1)}(a_k-a_\ell)\\
\Rightarrow \ \ a_k &= q^{-\alpha_k}\left( \frac{ q^{\alpha_k} - q^{-\alpha_k}
	}{ q-q^{-1} }\right) = q^{-\alpha_k}[\alpha_k]_q, \\
	\tilde{a}_k &= q^{-\overline{\alpha}_k}[\overline{\alpha}_k]_q.
\end{align*}
The above formulae allow for alternative expressions for the invariants
$\omega_k$, $\tilde{\omega}_k$, $\mu_r$, $\tilde{\mu}_r$ and so on.

We similarly have
$$
\tilde{a}_k - q^2\tilde{a}_r + q = q^{2(n-1)}(a_k - q^2a_r + q)
$$
so
$$
\tilde{a}_{0\ell} - q^2\tilde{a}_{0r} + q = q^{2(n-2)}(a_{0\ell}-q^2a_{0r} +
q).
$$
We also note that
\begin{align*}
	(\Lambda+\rho,\varepsilon_r - \varepsilon_k) &= \alpha_r-\alpha_k =
	\overline{\alpha}_r - \overline{\alpha}_k, \\
	(\Lambda+\rho\pm \varepsilon_r,\varepsilon_r - \varepsilon_k) &=
	\alpha_r-\alpha_k \pm 1=
	\overline{\alpha}_r - \overline{\alpha}_k\pm 1,
\end{align*}
and
\begin{align*}
[\alpha_r-\alpha_\ell - 1]_q &= q^{\alpha_r+\alpha_\ell-1}(q^2a_r - a_\ell -
	q)\\
	&= [\overline{\alpha}_r-\overline{\alpha}_\ell - 1]_q\\
	&= q^{\overline{\alpha}_r+\overline{\alpha}_\ell - 1}(q^2\tilde{a}_r -
	\tilde{a}_\ell - q),\\
[\alpha_r-\alpha_\ell + 1]_q &= q^{\alpha_r+\alpha_\ell-1}(a_r-q^2a_\ell+q)\\
	&= [\overline{\alpha}_r - \overline{\alpha}_\ell+1]_q 
= q^{\overline{\alpha}_r+\overline{\alpha}_\ell- 1}(\tilde{a}_r -
	q^2\tilde{a}_\ell+q),\\
[\alpha_r - \alpha_\ell]_q &= q^{\alpha_r+\alpha_\ell}(a_r-a_\ell)\\
	&= [\overline{\alpha}_r - \overline{\alpha}_\ell]_q =
	q^{\overline{\alpha}_r + \overline{\alpha}_\ell}(\tilde{a}_r -
	\tilde{a}_\ell).
\end{align*}
Furthermore, this implies
$$
\frac{ D_q[\Lambda\pm \varepsilon_r] }{ D_q[\Lambda] } = \prod_{\ell\neq r}
\frac{ [\alpha_r-\alpha_\ell\pm 1]_q }{ [\alpha_r-\alpha_\ell]_q }
$$
so that
\begin{align*}
	\frac{ D_q[\Lambda+ \varepsilon_r] }{ D_q[\Lambda] } &= \prod_{\ell\neq
	r}^n\frac{ q^{\alpha_r+\alpha_\ell-1}(a_r-q^2a_\ell+q) }{
		q^{\alpha_r+\alpha_\ell}(a_r-a_\ell) }\\
	&= \prod_{\ell\neq r}^n \frac{ q^{-1}a_r - qa_\ell+1 }{ a_r-a_\ell }\\
	&= \prod_{\ell\neq r}^n\frac{ q^{-1}\tilde{a}_r - q\tilde{a}_\ell+1 }{
		\tilde{a}_r - \tilde{a}_\ell },\\
	\frac{ D_q[\Lambda- \varepsilon_r] }{ D_q[\Lambda] } &= \prod_{\ell\neq
	r}^n\frac{ q^{\alpha_r+\alpha_\ell-1}(q^2a_r-a_\ell-q) }{
		q^{\alpha_r+\alpha_\ell}(a_r-a_\ell) }\\
	&= \prod_{\ell\neq r}^n \frac{ qa_r - q^{-1}a_\ell-1 }{ a_r-a_\ell }\\
	&= \prod_{\ell\neq r}^n\frac{ q\tilde{a}_r - q^{-1}\tilde{a}_\ell-1 }{
		\tilde{a}_r - \tilde{a}_\ell }.
\end{align*}
In summary:
\begin{align*}
[\alpha_r-\alpha_\ell]_q &= q^{\alpha_r+\alpha_\ell}(a_r-a_\ell) 
=[\overline{\alpha}_r-\overline{\alpha}_\ell]_q =
	q^{\overline{\alpha}_r+\overline{\alpha}_\ell}(\tilde{a}_r -
	\tilde{a}_\ell), \\
[\alpha_r-\alpha_\ell\pm 1]_q &= q^{\alpha_r+\alpha_\ell}(q^{\mp 1}a_r - q^{\pm
	1}a_\ell \pm 1)\\
	&= q^{\overline{\alpha}_r+\overline{\alpha}_\ell}(q^{\mp 1}\tilde{a}_r - q^{\pm
	1}\tilde{a}_\ell \pm 1),\\
\frac{ D_q[\Lambda\pm \varepsilon_r] }{ D_q[\Lambda] } 
	&= \prod_{\ell\neq r}\frac{ q^{\mp 1}{a}_r - q^{\pm
	1}{a}_\ell \pm 1 }{ {a}_r-{a}_\ell }\\
	&= \prod_{\ell\neq r}\frac{ q^{\mp 1}\tilde{a}_r - q^{\pm
	1}\tilde{a}_\ell \pm 1 }{ \tilde{a}_r-\tilde{a}_\ell }.
\end{align*}
Finally we also note that
\begin{align*}
	a_k - q^{-2}a_{0r}-q^{-1} &= \frac{ 1-q^{-2\alpha_k} }{ q-q^{-1} } -
	q^{-2}\frac{ 1-q^{-2\alpha_{0r}} }{ q-q^{-1} } - q^{-1}\\
	&= \frac{ q^{-2-2\alpha_{0r}} - q^{-2\alpha_k} }{ q-q^{-1} }\\
	&= q^{-(\alpha_k+\alpha_{0r}+1)}[\alpha_k - \alpha_{0r}-1]_q,
\end{align*}
and similarly
$$
\tilde{a}_k - q^2\tilde{a}_{0r}+q =
q^{-(\overline{\alpha}_{0r}+\overline{\alpha}_k-1)}[\overline{\alpha}_k -
\overline{\alpha}_{0r}+1]_q.
$$


\section*{Appendix E: Hopf $*$ structure (real $q>0$)}

Observe that we have a conjugation (or star) operation defined by
$$
e_i^\dagger = f_i,\ \ f_i^\dagger = e_i,\ \ h_i^\dagger = h_i,
$$
which extends to an (anti-linear) anti-homomorphism on all of $U_q(n)$ and thus
defines a conjugation (or star) operation of all of $U_q(n)$. Recall the
defining properties, $\forall a,b\in U_q(n),$ $\alpha,\beta\in\mathbb{C}$:
\begin{align*}
	(i) & (a^\dagger)^\dagger = a,\\
	(ii) & (ab)^\dagger = b^\dagger a^\dagger,\\
	(iii) & (\alpha a+\beta b)^\dagger = \overline{\alpha}a^\dagger +
	\overline{\beta}b^\dagger.
\end{align*}
With this operation, $U_q(n)$ is a Hopf $*$ algebra for generic real $q>0$,
i.e. $\Delta$, $\varepsilon$ determine $*$-algebra homomorphisms so
$$
\Delta(a)^\dagger = \Delta(a^\dagger),\ \ \varepsilon(a^\dagger) =
\overline{\varepsilon(a)},\ \ \forall a\in U_q(n).
$$

It is easily seen that for any $a\in U_q(n)$,
$$
S(a)^\dagger = S^{-1}(a^\dagger)
$$
(uniqueness of the antipode) - observe that $\tilde{S}(a) = S^{-1}(a)^\dagger$
is also an antipode. Moreover for the $R$-matrix we have
$$
R^\dagger = R^T,\ \ (R^T)^\dagger = R
$$
where $\dagger$ on the left hand side is the naturally induced conjugation
operation on $U_q(n)\otimes U_q(n)$.

Following the standard approach, we have \\
\underline{Definition}: A finite dimensional $U_q(n)$-module $V$ is called {\em
unitary} if it can be equipped with an inner product $\langle\ ,\ \rangle$ such
that
$$
\langle a^\dagger v,w\rangle = \langle v,aw \rangle,\ \ \forall a\in U_q(n),\
v,w\in V.
$$
\underline{Note}: If $\pi$ is the representation afforded by $V$ then this is
equivalent to
$$
\pi(a^\dagger)_{ji} = \overline{\pi(a)}_{ij}.
$$
That is, 
$$
\pi(a)^\dagger = \pi(a^\dagger).
$$

It is well known \cite{ZG1993} that all finite dimensional irreducible $U_q(n)$-modules
are equivalent to unitary ones. In particular the (undeformed) vector module is
unitary. However, care needs to be taken since the dual vector module $V^*$
defined in the usual way by
$$
\langle av^*,w \rangle = \langle v^*,S(a)w\rangle,\ \ v^*\in V^*,\ w\in V
$$
is {\em not} unitary (only equivalent to a unitary one). Indeed
\begin{align*}
\pi_0^*(a)^\dagger &= \pi_0^t(S(a))^\dagger \\
&= \pi_0^t(S(a)^\dagger) \\
	&= \pi_0^t(S^{-1}(a^\dagger))\\
	&= \pi_0^t(S(q^{2h_\rho}a^\dagger q^{-2h_\rho}))\\
	&= \pi_0^*(q^{2h_\rho}a^\dagger q^{-2h_\rho}).
\end{align*}
However the pseudo vector representation $\overline{\pi}_0$ is unitary. Hence
when discussing Wigner coefficients, matrix elements, etc, it is important to
work with unitary modules. Hence we assume that the representation afforded by
$V^*$ is given by $\overline{\pi}_0$ rather than $\pi_0^*$.

\subsection*{Fundamental matrices}

Since 
$$
R^\dagger = R^T,\ \ (R^T)^T = R
$$
it follows that 
$$
\tilde{A} = (q-q^{-1})^{-1}(\pi_0\otimes \id)(I - \tilde{R}^T\tilde{R})
$$
is self-adjoint, since $\pi_0$ is unitary. Thus
$$
\tilde{A}^\dagger = \tilde{A}.
$$
Writing
$$
\tilde{A} = \sum_{i,j} e_{ij}\otimes \tilde{A}_{ij}
$$
\begin{align*}
	\Rightarrow \ \ \sum_{i,j}e_{i,j}\otimes \tilde{A}_{ij} &= \tilde{A}\\
	&= \tilde{A}^\dagger = \sum_{j,i}e_{ji}\otimes
	(\tilde{A}_{ij})^\dagger\\
	\Rightarrow \ \ (\tilde{A}_{ij})^\dagger &= \tilde{A}_{ji}.
\end{align*}
Similarly, since $\overline{\pi}_0$ is unitary,
$$
(A_{ij})^\dagger = A_{ji}.
$$

%
%
%

%

\end{document}